\documentclass[aps,pre,psfig,twocolumn,showpacs,superscriptaddress,preprin,floatfixt]{revtex4-1}
\usepackage{amsmath}
\usepackage{mathtools}
\usepackage{amssymb}
\usepackage{color}
\usepackage{graphics}
\usepackage{epsfig}
\usepackage[normalem]{ulem} 
\usepackage{cancel}
\usepackage[mathlines]{lineno}
\graphicspath{{Figures/}}
\usepackage{xtab,afterpage,longtable}
\usepackage{booktabs}   
\usepackage{ltablex}
\relax
\usepackage{float}

\begin{document}
\title{Bispectral analysis of nonlinear mixing in a periodically driven Korteweg-de Vries system}
\author{Ajaz Mir}
\email{2018rph0028@iitjammu.ac.in}
\affiliation{Indian Institute of Technology Jammu, Jammu, J\&K,  181221, India}
\author{Sanat Tiwari}
\email{sanat.tiwari@iitjammu.ac.in}
\affiliation{Indian Institute of Technology Jammu, Jammu, J\&K,  181221, India}
\author{Abhijit Sen}
\affiliation{Institute for Plasma Research, Gandhinagar, Gujarat, 382428, India}
\date{\today}
\begin{abstract}
\textcolor{black}{The nonlinear response of a periodically driven Korteweg-de Vries model system is studied using a variety of nonlinear drivers and compared to previous results obtained for a purely time-dependent sinusoidal driver [Phys. Plasmas \textbf{27}, 113701 (2020)]. It is found that a nonlinear driver in the form of a cnoidal square wave or a travelling wave driver produces a spectral response that is closer to experimental observations [Phys. Rev. Lett. \textbf{92}, 085001 (2004)] than that predicted by the simple sinusoidal driver. 
Using a bispectral analysis, we also firmly establish that the nature of the nonlinear oscillations, due to the interaction between the periodic source and the inherent collective mode of the system, is predominantly governed by a three-wave mixing process. Furthermore, by studying the variation in the mixing pattern, from a broad to a sparse frequency spectrum,  as a function of the driver frequency and its functional form, we propose a  means of tailoring the nature of such patterns.
Our results could find useful applications in the experimental interpretation and manipulation of nonlinear wave mixing patterns in weakly nonlinear and dispersive plasma systems or similar phenomena in neutral fluids.}
\end{abstract}
\maketitle
\section{Introduction}
\label{intro_NLM}
\paragraph*{}
The Korteweg-de Vries (KdV) model equation has served as a standard paradigm for the description of many low-frequency nonlinear phenomena in various nonlinear dispersive media such as neutral fluids, plasmas, optical fibres, vibrating lattices \textit{etc.}~\citep{Washimi_PRL_1966,Gasch_PRA_1986, Hao_PRA_2001, Leblond_PRA_2008, Peradze_LTP_2019, Redor_PRL_2019, Congy_PRA_2016, Kamchatnov_PRA_2004}. As a fully integrable partial differential equation, the KdV equation admits a variety of exact solutions, including solitons, cnoidal waves, dispersive shocks \textit{etc.} that have been used to describe experimental observations of nonlinear wave phenomena in the aforementioned systems. 
\textcolor{black}{In plasmas, the KdV equation has successfully modelled the existence and evolution of  ion and dust acoustic shocks as well as solitary waves~\citep{Mo_PRL_2013, Bandyopadhyay_PRL_2008, Nakamura_PRL_1984}. More recently, a forced  KdV (fKdV) model was used to explain the emission of precursor magneto-sonic solitons, created by a charge bunch moving in a magnetized plasma~\citep{Atul_NJP_2020}.}
\paragraph*{}
The KdV model has also been investigated in the presence of an external time and space varying perturbation in order to investigate the response of a driven nonlinear medium. Such a forced KdV equation that had a travelling wave source was used to determine the conditions for the resonant excitation of nonlinear oscillatory waves in a dispersive medium by Vainberg {\it et al.}~\citep{Vainberg_RPQE_1983}. Using a chirped frequency driving perturbation, Friedland {\it et al.}~\citep{Friedland_PRE_2015} studied the anomalous autoresonance threshold for the excitation of large amplitude travelling waves in a  KdV model system. Aranson {\it et al.}~\citep{Aranson_PRA_1992} took a similar approach of chirping the frequency of the driver to excite solitons in the system. \textcolor{black}{In particular, for plasmas,  the fKdV model predicts existence of precursor and pinned solitons~\citep{Garima_PRE_2021,Surabhi_PRE_2016}. These nonlinear emissions can prove useful in space awareness applications, such as in the indirect detection of charged space debris orbiting in the earth's ionosphere~\citep{Sen_ASR_2015}.} More recently, \textcolor{black}{ Mir {\it et al.}~\cite{Ajaz_POP_2020} used the fKdV model to study nonlinear wave mixing in a dusty plasma system and related their findings to past experimental observations of Nosenko {\it et al.}~\citep{Nosenko_PRL_2004}}. The form of the driver, in their case, was chosen to be a simple time-varying sinusoidal function for which an exact analytic solution of the model could be obtained. Using such an exact solution, time series data of the system response was obtained for parameters close to the experimental conditions and the power spectrum of this model data was compared with the experimental power spectrum. Nonlinear mixing (NLM) was revealed by the presence of the combination of frequencies of various harmonics of the inherent mode of the system and the driving wave.
\paragraph*{}
While such identification of combination frequencies provides quick first-hand evidence of nonlinear wave mixing, it does not conclusively establish the origin of the combination due to a three-wave mixing process. This is because the power spectrum does not explicitly provide the phase coupling information about the interactions. A more precise tool for establishing the physical origin of the wave mixing is a bispectral analysis~\citep{Rao_Gabr_SV_1984, Nikias_IEEE_1987, Kim_IEEE_1979} that looks at the triple-correlation of the time series of any dynamical quantity. A finite correlation is obtained for a frequency triad $F(f_1)$, $F(f_2)$ and $F(f_1+f_2)$ when they are formed by a coherent phase coupling mechanism (where $F$ is the discrete Fourier component at a given frequency). 
There will be no correlation among frequencies if they are spontaneously excited modes, i.e., no coherent phase coupling is involved. Bispectral analysis has been extensively used in the past to investigate coherent nonlinear interactions in plasmas~\citep{Milligen_PRL_1995, Kim_PRL_1997, Nosenko_PRE_2006} as well as in many biomedical~\citep{Siu_IEEE_2008,Tacchino_IEEE_2020} and engineering applications~\citep{Hillis_PRCA_2006,Hall_IEEE_1995}.
\paragraph*{}
In this paper, we extend the earlier spectral analysis of Mir {\it et al.}~\cite{Ajaz_POP_2020} and carry out a bispectral analysis to firmly establish the nature of the nonlinear mixing phenomenon. We also expand the scope of their model calculations by going beyond the simple time-varying sinusoidal driver to examine the influence of different functional forms of the driver on the mixing process. 
\textcolor{black} {These different functional forms of the driver can arise in a variety of situations. For example, in a typical experimental situation, if the driving wave is one of the normal modes of the system, then it is likely to be of a nonlinear form, {\it e.g.} ion acoustic waves or dust acoustic waves (DAWs) have been observed to grow nonlinearly into cnoidal or cnoidal square waveforms in many experiments 
~\citep{Flanagan_POP_2010,Suranga_PRE_2012,Neeraj_POP_2015,Teng_PRL_2009}. 
The presence and evolution of nonlinear modes or disturbances are, in fact, generic to nonlinear media such as cnoidal waves in metamaterials~\citep{Mo_PRE_2019}, cnoidal-square waveforms of dust acoustic waves~\citep{Liu_POP_2018}, and sawtooth-like shock waves~\citep{Heinrich_PRL_2009}. When the driving wave is applied externally, it can also take a variety of waveforms,  such as using a cnoidal waveform electrical or  optical signals to generate frequency combs in  Kerr microresonators ~\citep{Kholmyansky_PRA_2019}, use of a sawtooth wave for higher harmonic generation in a plasma~\citep{Sharma_PSST_2020} and the use of cnoidal waves to generate nonlinear frequency combs in microring resonators~\citep{Qi_JOSAB_2017}. The present work is aimed at studying the impact of such waveforms on the nonlinear mixing phenomena in plasmas.}
\paragraph*{}
It is seen that changing the profile and functional dependence of the driver can significantly alter the response pattern of the system. In particular, a travelling wave source is seen to excite additional frequencies that are not seen in response to purely time-varying sources.This enhanced spectrum is shown to bear a closer resemblance to the \textcolor{black}{dusty plasma} experimental data of Nosenko {\it et al.}~\citep{Nosenko_PRL_2004} than the model system studied by Mir {\it et al.}~\citep{Ajaz_POP_2020}. In addition to changing the functional form of the driver, we also examine the effect of changing the driver frequency relative to the natural frequency of the non-driven system. We find a significant difference in the resultant spectral patterns depending upon whether the driver frequency is larger or smaller than the natural frequency. This asymmetry in the response pattern is explained based on a higher-order mixing process. Our findings thereby offer a means of tailoring the response patterns of driven systems that may find practical applications.
\paragraph*{}
The paper is organized as follows. 
The model fKdV equations with different forms of  purely  time-dependent forcing terms $F_s(t)$ or  with a  spatio-temporal forcing term $F_s(x,t)$  are described in section~\ref{fKdV_model}.
The effect of different forcing forms on nonlinear mixing is first discussed in section~\ref{NLM_fKdV} in terms of the differences in the power spectral densities (PSD) of the time series data of the numerical solutions of the fKdV model. This is followed up in section~\ref{NLM_BI_fKdV} by a bispectral analysis of the same data and its physical interpretation. Section ~\ref{NLM_tunable} is devoted to a discussion on the effect of tunable drivers on the NLM process and a discussion on the physical origin of the asymmetry in the driven response. A brief discussion and some concluding remarks are given in section~\ref{summary}.
\section{The forced Korteweg-de Vries model}
\label{fKdV_model}
\paragraph*{}
A generalized form of the fKdV equation for the perturbed density, $n(x,t)$ can be written down as \citep{Ajaz_POP_2020}:
\begin{equation}
  \frac{\partial n(x,t)}{\partial t} 
  +\ 
  \alpha\  n(x,t)\frac{\partial n(x,t)}{\partial x} 
  +\ 
  \beta \frac{\partial^3 n(x,t)}{\partial x^3} 
  =\ 
  F_s(x,t).
\label{fKdV_eqn}
\end{equation}
\paragraph*{}
\textcolor{black}{The Eq.~(\ref{fKdV_eqn})} can be derived from the full set of cold fluid equations for the dust component in the weakly nonlinear and weakly dispersive regime \cite{Sen_ASR_2015}. The ions and electrons, as lighter species compared to the dust, are assumed to obey Boltzmann relations. $\alpha$ and $\beta$ are the characteristic parameters quantifying nonlinearity and dispersion of the medium, respectively.  $F_s(x,t)$ is the external spatio-temporal forcing. For $F_s(x,t)=0$ one recovers the standard KdV equation that admits exact nonlinear solutions in the form of solitons~~\citep{Farina_PRL_2001, Bandyopadhyay_PRL_2008, Mo_PRL_2013,Peradze_LTP_2019} and periodic cnoidal waves~\citep{Mahmood_POP_2014, Liu_POP_2018}.
\textcolor{black}{ In the context of dusty plasmas, Liu {\it et al.}~\citep{Liu_POP_2018} have shown that a cnoidal wave solution provides an excellent fit to their experimental observations of self-excited dust acoustic waves sustained in an RF plasma~\citep{Flanagan_POP_2011}.}
\subsection{Korteweg-de Vries model with a time-dependent forcing, $F_s(t)$}
\label{fKdV_time}
\paragraph*{}
An analytic solution of Eq.~(\ref{fKdV_eqn}) with a purely time-dependent sinusoidal forcing that is based on  Hirota's approach~\citep{Salas_NLA_2011} has been presented earlier in Ref. \cite{Ajaz_POP_2020} and is given by:
\begin{eqnarray}
&& n(x, t) =\ 
\phi(t)
+\ \mu \ cn^2 \bigg[
\frac{\sqrt{\mu \alpha}}{2 \sqrt{\beta \kappa(\kappa +2)}} \xi(x,t) ;\ \kappa  
\bigg] \nonumber \\
&& \xi(x,t) = \bigg( x  - \frac{\kappa + \kappa^2 - 1}{\kappa(\kappa + 2)} \alpha \mu t - \alpha \psi(t)  \bigg).
\label{fKdV_analytic_solution}
\end{eqnarray}
In Eq.~(\ref{fKdV_analytic_solution}), $cn$ is the Jacobi elliptic function, $ \mu$ is the amplitude of the cnoidal wave, and $\kappa$ is the elliptic modulus that quantifies the nonlinear nature of the cnoidal wave. 
\textcolor{black}{  In the context of dusty plasma, the elliptic modulus $\kappa$ and the total harmonic distortion of DAW~\citep{Flanagan_POP_2010} were used by Liu {\it et al.}~\citep{Liu_POP_2018}, to quantify the nonlinearity of their experimentally observed waveforms.}
With $\kappa \rightarrow 0$, the wave attains a cosine waveform, and as $\kappa \rightarrow 1$, the waveform becomes nonlinear with the cosine wave getting converted to a cnoidal waveform. At $\kappa = 1$ when $\text{cn}^2(x,t ;\ \kappa) = \text{sech}^2(x,t)$~\citep{Abramowitz_Dover_1965}, the cnoidal wave takes the form of a single soliton.
Furthermore, the quantities $\phi$ and $\psi$ are defined as,
\begin{equation}
\phi(t) =\ \int  F_s(t)\ dt \hspace{0.6cm} \text{and} \hspace{0.6cm} \psi(t) =\ \int \phi(t)\ dt .
\label{phi_psi}
\end{equation}
\begin{table}[ht!]
\caption{Functional \textcolor{black}{form} of different forcing profiles}
{\renewcommand{\arraystretch}{1.75}
\begin{tabular}{|c|c|}
\hline
{ \bf Forcing profiles } & $ F_s(t)$ \textcolor{black}{or $F_s(x,t)$}  
\\ \hline
\shortstack{ \\ Sinusoidal  \\ $ F_{\sin}(t)$}   &  $A_s \sin(2 \pi f_2 t)$     
\\ \hline  
\shortstack{  \\ Cnoidal  \\ $F_{cn}(t)$ }     &  $ A_s cn[4 K(\kappa_s) f_2 t; \kappa_s]$     
\\ \hline
\shortstack{ \\Cnoidal-square  \\ $F_{cn^2}(t)$ } &  
\shortstack{ \\$ A_s cn^2[2 K(\kappa_s) f_2 t; \kappa_s]$  
\\ $F_{cn^2} (t) = F_{cn^2} (t)  - \left\langle F_{cn^2} (t) \right\rangle$}   
\\ \hline  
\shortstack{\\ \textcolor{black}{Travelling wave} \\  \textcolor{black}{$F_s(x,t)$} }  & 
\shortstack{ \\  $A_s \sin (k_s x - 2\pi f_2 t )$ \\ \textcolor{black}{Solving fKdV numerically} }
\\ \hline 
\end{tabular}
\label{Tab_1}
}
\end{table}
The $\phi(t)$ and $\psi(t)$ can be calculated either analytically or numerically depending on the choice of the forcing profile. 
As a generalization of the work reported in Ref.~\cite{Ajaz_POP_2020}, where only a sinusoidal form of $F_s(t)$ was used, we have considered a variety of functional forms for the drivers in our present work. 
These different forcing terms are listed in Table~\ref{Tab_1}.  For the case of a sinusoidal forcing profile, $\phi(t)= - \left( A_s/ (2\pi f_2) \right) \cos(2\pi f_2 t)$ and $\psi(t) = -\left(A_s/ (2\pi f_2)^2 \right) \sin(2\pi f_2 t)$. Here $f_2$, $A_s$ and $\kappa_s$ are the forcing frequency, forcing amplitude and the forcing elliptic modulus corresponding to each forcing form $F_s(t)$ respectively.
\begin{figure} [ht!]
\includegraphics[width = \columnwidth]{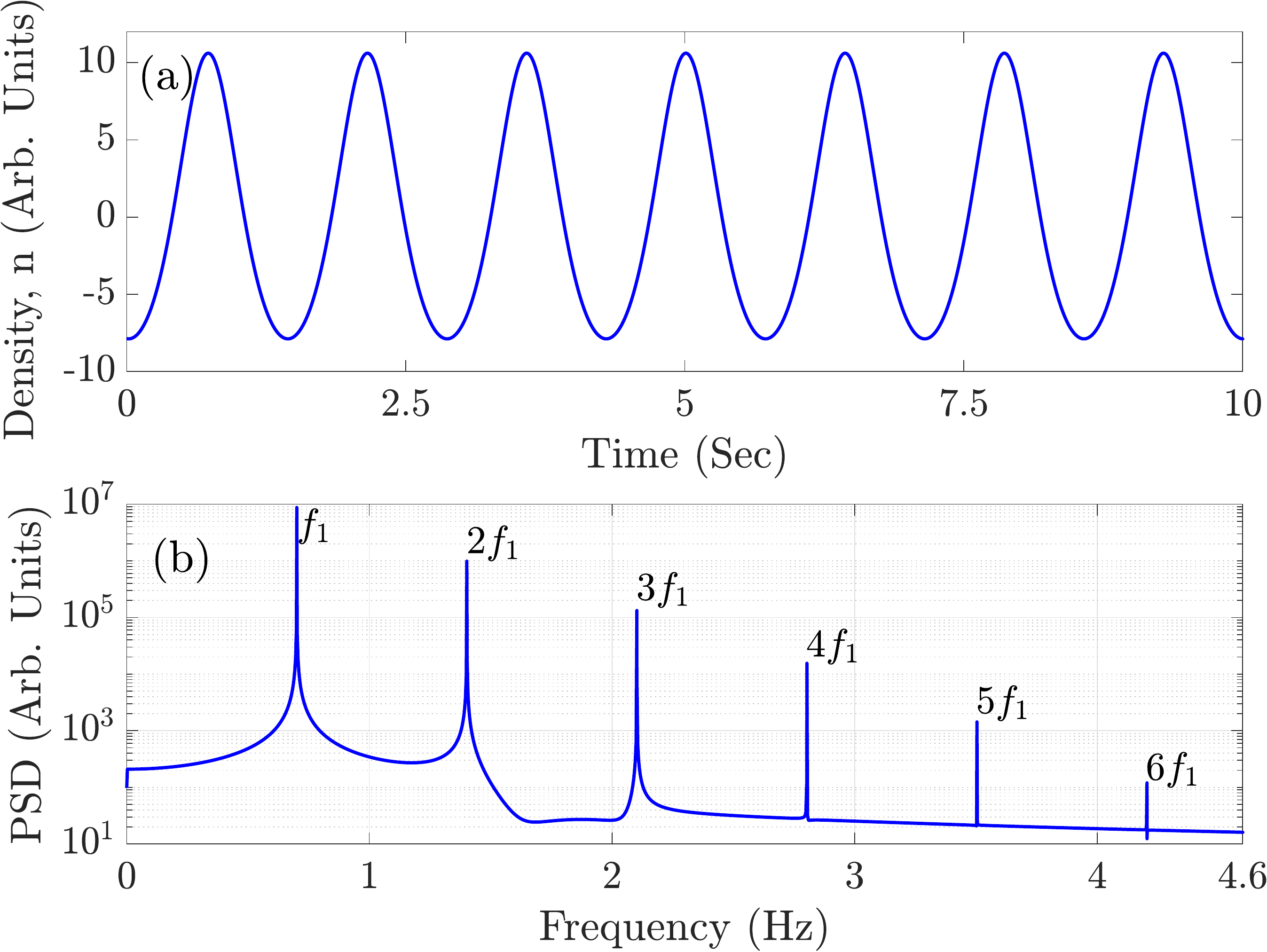}
 \caption{ 
(a) Time series of KdV for $\mu = 18.5$, $\kappa =0.7$ (such that $f_1=0.7$ Hz) with $\alpha = \beta = 1$. 
(b) Power spectrum of (a). 
The nonlinear wave (non-sinusoidal) nature is evident from both the time series (a) and the presence of harmonics in its PSD (b). 
}
\label{Fig_1}
\end{figure}
\paragraph*{}
A truncated Fourier series expansion can approximate the non-sinusoidal forcing profiles. 
\textcolor{black}{ In the context of dusty plasma, Merlino {\it et al.}~\citep{Merlino_POP_2012} obtained excellent fit to their experimental DAW profiles by retaining terms up to the second harmonic in the Fourier series expansion of the square of the cnoidal function, namely,} 
\begin{eqnarray}
\nonumber
F_s (t) =\ F_{cn^2}(t) = A_s cn^2[2 K(\kappa_s) f_2 t; \kappa_s]
\\
\nonumber
F_{cn^2} (t) \approx (A_s/2)\{ \cos(2\pi f_2 t) \\
\nonumber
          +\ 0.02A_s \cos(2\pi (2f_2) t) \\
          +\ 0.004A_s \cos(2\pi (3f_2) t) \}.
\label{CN2_Fourier_series}
\end{eqnarray}
Such a Fourier series representation of a non-sinusoidal forcing term allows one to solve the fKdV system analytically by using the general solution described by \eqref{fKdV_analytic_solution} and (\ref{phi_psi}). This facilitates the study of the nonlinear mixing phenomenon as a function of the various forms of the driver. 
\paragraph*{}
A nonlinear cnoidal time series solution of KdV equation and its PSD are shown in Fig.~\ref{Fig_1}(a) and Fig.~\ref{Fig_1}(b) respectively. The PSD shows a fundamental frequency at $f_1=0.7$ Hz along with the even and odd harmonics of the wave at $2f_1$, $3f_1$ and so on. 
\begin{figure*}[ht!]
\includegraphics[width = 0.9\textwidth]{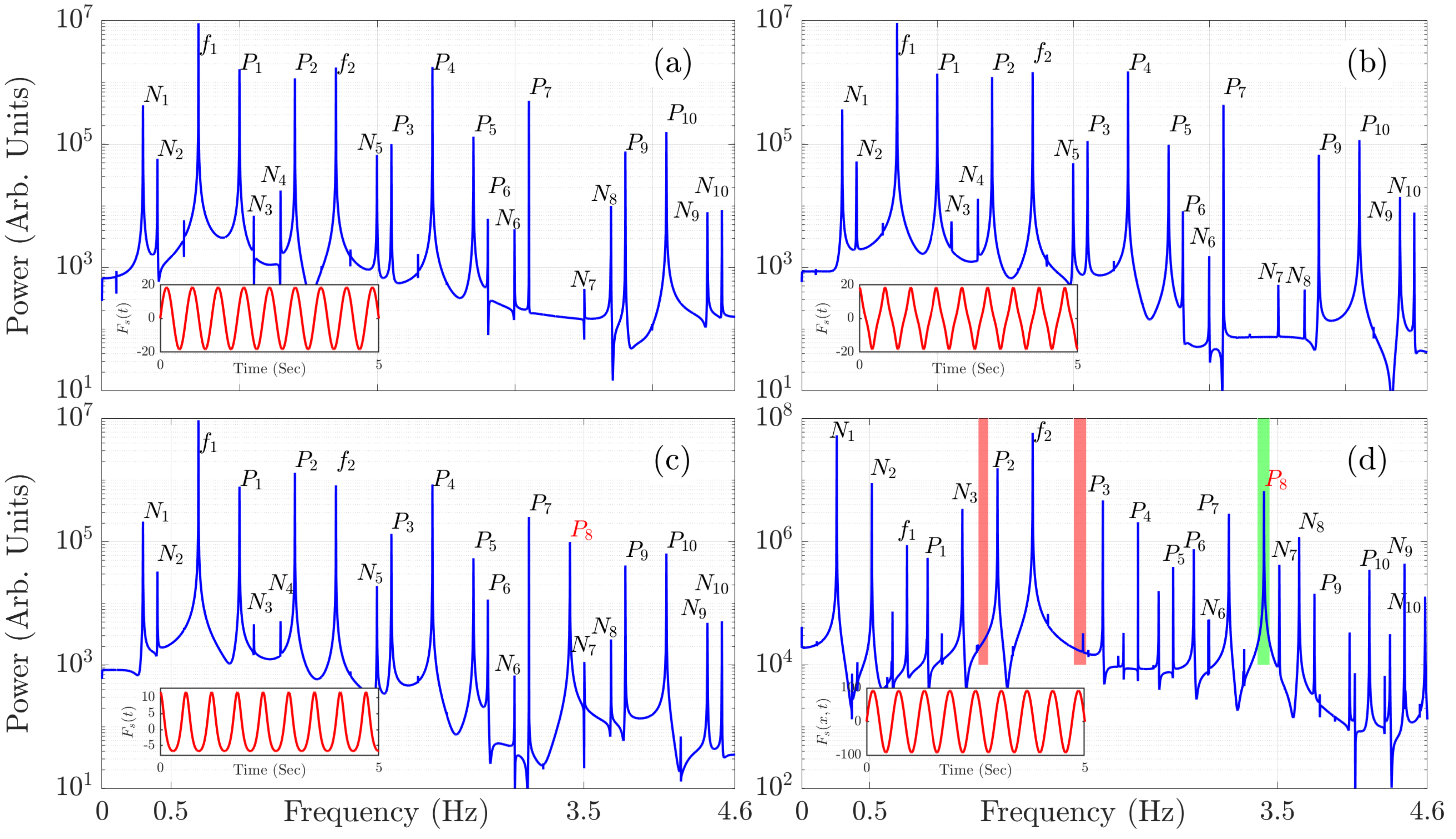}
\caption{
Power spectrum of time series obtained from fKdV model for 
(a) sinusoidal forcing $ F_s(t) = F_{\sin}(t) = A_s \sin(2\pi f_2 t)$ , 
(b) cnoidal wave forcing $F_s(t) = F_{cn}(t) = A_s cn[4 K(\kappa_s) f_2 t; \kappa_s]$, 
(c) cnoidal-square wave forcing $F_s(t) = F_{cn^2} (t) = A_s cn^2[2 K(\kappa_s) f_2 t; \kappa_s]$.
Here $A_s = \mu$, $\kappa_s = 0.9$ and $f_2=1.7$ Hz.
(d) Travelling wave forcing $ F_s(x,t) = A_s \sin(k_s x - 2\pi f_2 t)$ with no initial condition, i.e., $n(x,0) = 0$. Here $A_s = 5\mu $, $k_s = 9 k_0$ with $k_0 = (2\pi)/ L$, which corresponds to frequency $f_1 = k_s^3/(2\pi) = 0.7$ Hz for a system of length $L=11\pi$ and $f_2 = 1.7$ Hz. Also $\alpha =\ \beta =\ 1$ in each case. 
}
\label{Fig_2}
\end{figure*}
\begin{table}[ht!]
\caption{Dominant frequencies observed in the fKdV model for various forcing forms as shown in Fig.~\ref{Fig_2}.}
{\renewcommand{\arraystretch}{1.25}
\begin{tabular}{|l|l|l|l|l|}
\hline   
  \hspace{0.5cm}   &  Fig.~\ref{Fig_2}(a)  \hspace{0.1cm}  &  Fig.~\ref{Fig_2}(b)  \hspace{0.1cm} 
  & Fig.~\ref{Fig_2}(c)  \hspace{0.1cm}   & Fig.~\ref{Fig_2}(d) \hspace{0.1cm}  \\ \hline
Frequency~(Hz)     & 
\shortstack{\\ $F_{\sin}(t)$ \\ $A_s = \mu$}                    & 
\shortstack{\\ $F_{cn}(t)$ \\ $A_s = \mu$ \\ \ $\kappa_s = 0.9$}  & 
\shortstack{\\ $F_{cn^2}(t)$ \\ $A_s = \mu$ \\ \ $\kappa_s = 0.9$}  & 
\shortstack{\\ \textcolor{black}{ $F_s(x,t)$} \\ \textcolor{black}{$A_s = 5\mu$} \\ \ \textcolor{black}{$k_s = 9 k_0$}
   }\\ \hline
 $f_1=0.7$            & \checkmark   & \checkmark   &  \checkmark  &  \checkmark   \\ \hline
 $f_2=1.7$            & \checkmark   & \checkmark   &  \checkmark  &  \checkmark   \\ \hline 
 $P_1 =f_2 - f_1$     & \checkmark   & \checkmark   &  \checkmark  &  \checkmark   \\ \hline  
 $P_2=2f_1$           & \checkmark   & \checkmark   &  \checkmark  &  \checkmark   \\ \hline 
 $P_3=3f_1$           & \checkmark   & \checkmark   &  \checkmark  &  \checkmark   \\ \hline 
 $P_4=f_1+f_2$        & \checkmark   & \checkmark   &  \checkmark  &  \checkmark   \\ \hline 
 $P_5=2f_2 - f_1$     & \checkmark   & \checkmark   &  \checkmark  &  \checkmark   \\ \hline 
 $P_6=4f_1$           & \checkmark   & \checkmark   &  \checkmark  &  \checkmark   \\ \hline 
 $P_7=2f_1 + f_2$     & \checkmark   & \checkmark   &  \checkmark  &  \checkmark   \\ \hline
 $P_8=2f_2 $          &              &              & \checkmark   & \checkmark   \\ \hline
 $P_9=3f_1 + f_2$     & \checkmark   & \checkmark   &  \checkmark  &  \checkmark   \\ \hline
 $P_{10} =2f_2 + f_1$ & \checkmark   & \checkmark   &  \checkmark  &  \checkmark   \\ \hline
 $N_1=f_2-2f_1$       & \checkmark   & \checkmark   &  \checkmark  &  \checkmark  \\ \hline 
 $N_2=3f_1-f_2$       & \checkmark   & \checkmark   &  \checkmark  &  \checkmark  \\ \hline
 $N_3=4f_1-f_2$       & \checkmark   & \checkmark   &  \checkmark  &  \checkmark  \\ \hline 
 $N_4=2f_2-3f_1$      & \checkmark   & \checkmark   &  \checkmark  &    \\ \hline
 $N_5=2(f_2-f_1)$     & \checkmark   & \checkmark   &  \checkmark  &    \\ \hline
 $N_6=3(f_2-f_1)$     & \checkmark   & \checkmark   &  \checkmark  &  \checkmark  \\ \hline 
 $N_7=5f_1$           & \checkmark   & \checkmark   &  \checkmark  &  \checkmark  \\ \hline   
 $N_8=3f_2-2f_1$      & \checkmark   & \checkmark   &  \checkmark  &  \checkmark  \\ \hline   
 $N_9=3f_2-f_1$       & \checkmark   & \checkmark   &  \checkmark  &  \checkmark  \\ \hline 
 $N_{10}=4f_1+f_2$    & \checkmark   & \checkmark   &  \checkmark  &  \checkmark  \\ \hline 
\end{tabular}
}
\label{Tab_2}
\end{table}
\subsection{Korteweg-de Vries model with spatio-temporal forcing, $F_s(x,t)$}
\label{fKdV_spacetime}
\paragraph*{}
The KdV equation with a spatio-temporal travelling waveform of the forcing is given by:
\textcolor{black}{
\begin{eqnarray}
\nonumber
  \frac{\partial n(x,t)}{\partial t} 
  +\ 
  \alpha\  n(x,t)\frac{\partial n(x,t)}{\partial x} 
  +\ 
  \beta \frac{\partial^3 n(x,t)}{\partial x^3} 
  \\
  =\ 
  A_s \sin(k_s x -\ 2\pi f_2 t) .
\label{fKdV_travel}
\end{eqnarray}
}
We have solved  Eq.~(\ref{fKdV_travel}) numerically using a finite difference scheme. The code has been validated by reproducing the results of Sen \textit{et al.}~\citep{Sen_ASR_2015}. 
The frequency corresponding to the forcing wave-vector $k_s$ is chosen to satisfy the  linear dispersion relation obtained by setting $\alpha=0$ in the standard KdV equation, namely  $f_1 = -\beta k_s^3/(2\pi)$. Here, $k_s = n k_0$ with $k_0 = 2\pi / L$ being the minimum wave-vector associated with a system of length $L$. 
For the numerical solution of  Eq.~(\ref{fKdV_travel}) we have taken an initial condition of  $n(x,0)=0$. 
\begin{figure*}[ht!]
\includegraphics[width = 0.99\textwidth]{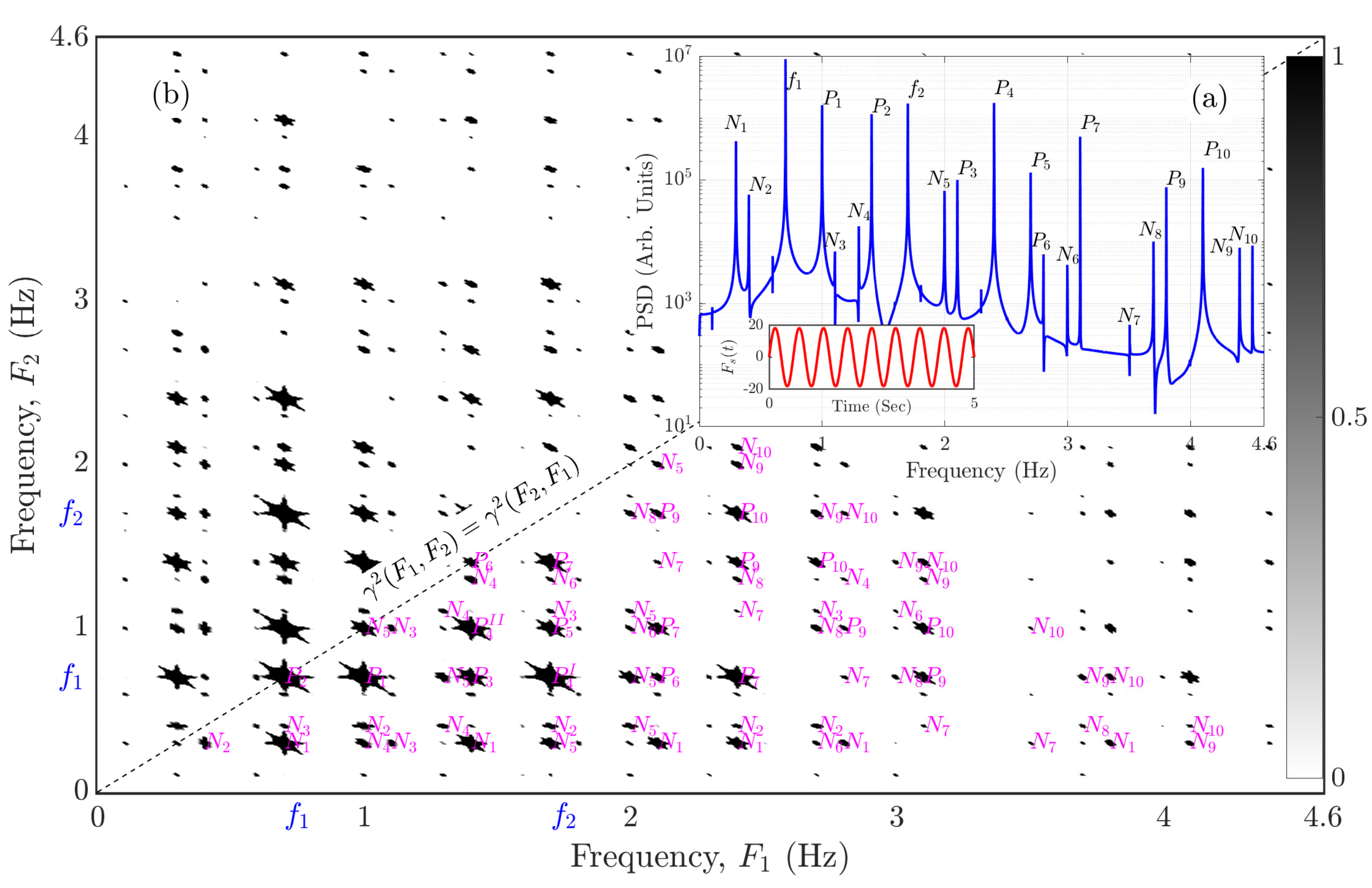}
\caption{
PSD (a) and bicoherence (b) of fKdV for sinusoidal forcing. 
Inset (a) is the PSD of fKdV for $F_{s}(t) = F_{\sin}(t) = A_s\sin(2\pi f_2 t )$ with $A_s =18.5$ and $f_2=1.7$ Hz. Initial parameters are $\mu = 18.5$, $\kappa =0.7$ (such that $f_1 =0.7$ Hz) and  $\alpha = \beta = 1$. 
The small inset within (a) shows the form of sinusoidal forcing profile.
The bicoherence map shows patches $(\gamma^2 \approx 1)$ indicating that the waves at $F(f_1)$, $F(f_2)$ and $F(f_1)+F(f_2)$ are not only frequency coupled but are phase coupled as well. This confirms coherent nonlinear interaction between the waves at $F(f_1)$ and $F(f_2)$.
}
\label{Fig_3}
\end{figure*}
\section{Nonlinear mixing in the fKdV model with different forcing forms}
\label{NLM_fKdV}
\paragraph*{}
To investigate the NLM process under various non-sinusoidal time-dependent forcing forms, we have used  semi-analytic solutions by  using  \eqref{fKdV_analytic_solution} and (\ref{phi_psi}). Exact analytic forms are possible for cases when $\phi$  and $\psi$ in (\ref{phi_psi}) are exactly integrable. Sinusoidal forcing is one such example. 
While the solutions apply to all nonlinearity regimes and dispersion, we have taken one parameter set throughout the paper for uniformity. We have used $\mu=18.5$, $\kappa=0.7$ (such that $f_1=0.7$ Hz) and $\alpha = \beta =1$ for all cases in the manuscript unless specified otherwise.
\paragraph*{}
The frequencies in the power spectrum as displayed in Fig.~\ref{Fig_2} are for (a) sinusoidal, (b) cnoidal wave, and (c) cnoidal-square wave driver in the fKdV equation. Their values are tabulated in Table~\ref{Tab_2}. All listed frequencies match the sum and difference frequencies of $f_1$ (natural KdV mode), $f_2$ (fundamental forcing mode) and their harmonics. These additional modes are generated via the three-wave mixing mechanism, which is further confirmed via a bispectral analysis described in section~\ref{NLM_BI_fKdV}.
\paragraph*{}
The spectrum shown in Fig.~\ref{Fig_2} for each non-sinusoidal forcing form is qualitatively similar to the sinusoidal forcing case in the range of frequencies that carry significant power. A noticeable difference is for the cnoidal-square forcing case where we obtain an additional frequency $P_8 = 2f_2$ that also exists in the power spectrum of the fluctuations measured in the dusty plasma experiment of Nosenko {\it et al.}~\citep{Nosenko_PRL_2004}. 
However, this could be because the cnoidal-square forcing form inherently contains the frequency mode at $P_8 = 2f_2$ in contrast to other forcing forms, as can be seen in Table~\ref{Tab_2}.
\paragraph*{}
We have also observed NLM in fKdV model with a travelling wave forcing form $F_s(x,t) =  A_s \sin(k_s x - 2\pi f_2 t)$.  The spectrum due to travelling wave forcing is shown in Fig.~\ref{Fig_2}(d) and is also tabulated in Table~\ref{Tab_2}. The spectrum obtained for this case is unique compared to all time-dependent forcing cases in two ways. First, certain frequencies are missing in the spectrum and are named as $N_4$ and $N_5$ in Table~\ref{Tab_2}. The same has been shown as red bands in Fig.~\ref{Fig_2}(d). Second, 
we have also observed a frequency peak $P_8 = 2 f_2$ marked in green patch in Fig.~\ref{Fig_2}(d).
Both the missing and the extra frequency are consistent with the observations in the dusty plasma experiment~\citep{Nosenko_PRL_2004}. In this case, the extra frequency is not due to the inherent property of the driver but is a genuine outcome of an NLM process.
\section{Bispectral analysis of nonlinear mixing}
\label{NLM_BI_fKdV}
\paragraph*{}
The bispectral analysis is a statistical tool that provides a quantitative measure of a  coherent nonlinear interaction process. If a coherent nonlinear interaction exists between three oscillations at frequencies $F_1$, $F_2$ and $F_1+F_2$, a peak will be generated in the principal domain of the bispectrum at the intersection between $F_1$ and $F_2$. The bispectrum of a dynamic process is always a complex quantity, and is defined by~\citep{Rao_Gabr_SV_1984,Kim_IEEE_1979,Nikias_IEEE_1987,Raju_PPCF_2003, Siu_IEEE_2008}
\begin{equation}
\label{def_bispectrum}
B(F_1, F_2) =\ \langle F(f_1) F(f_2) F^{\star}(f_1+f_2) \rangle
\end{equation}
where $\langle ... \rangle$ is the ensemble average over multiple samples, $F$ is the Fourier transform, $\star$ is the complex conjugate and $f_1$, $f_2$ are the two frequencies of the triad ($F_1$, $F_2$, $F_1+F_2$). The bispectrum $B(F_1, F_2)$ is a function of two frequencies and is a non-zero quantity only if a phase coupling exists between the frequency triplet $F_1$, $F_2$ and $F_1+F_2$. $B(F_1, F_2)$ is identically zero for spontaneously generated modes, i.e., the modes generated without phase coupling.  The bispectrum's ability to retain the phase information lost by the power spectrum (the Fourier transform of a signal) makes it a useful tool for analyzing coherent nonlinear interactions.
\begin{figure*}[ht!]
\includegraphics[width = 0.99\textwidth]{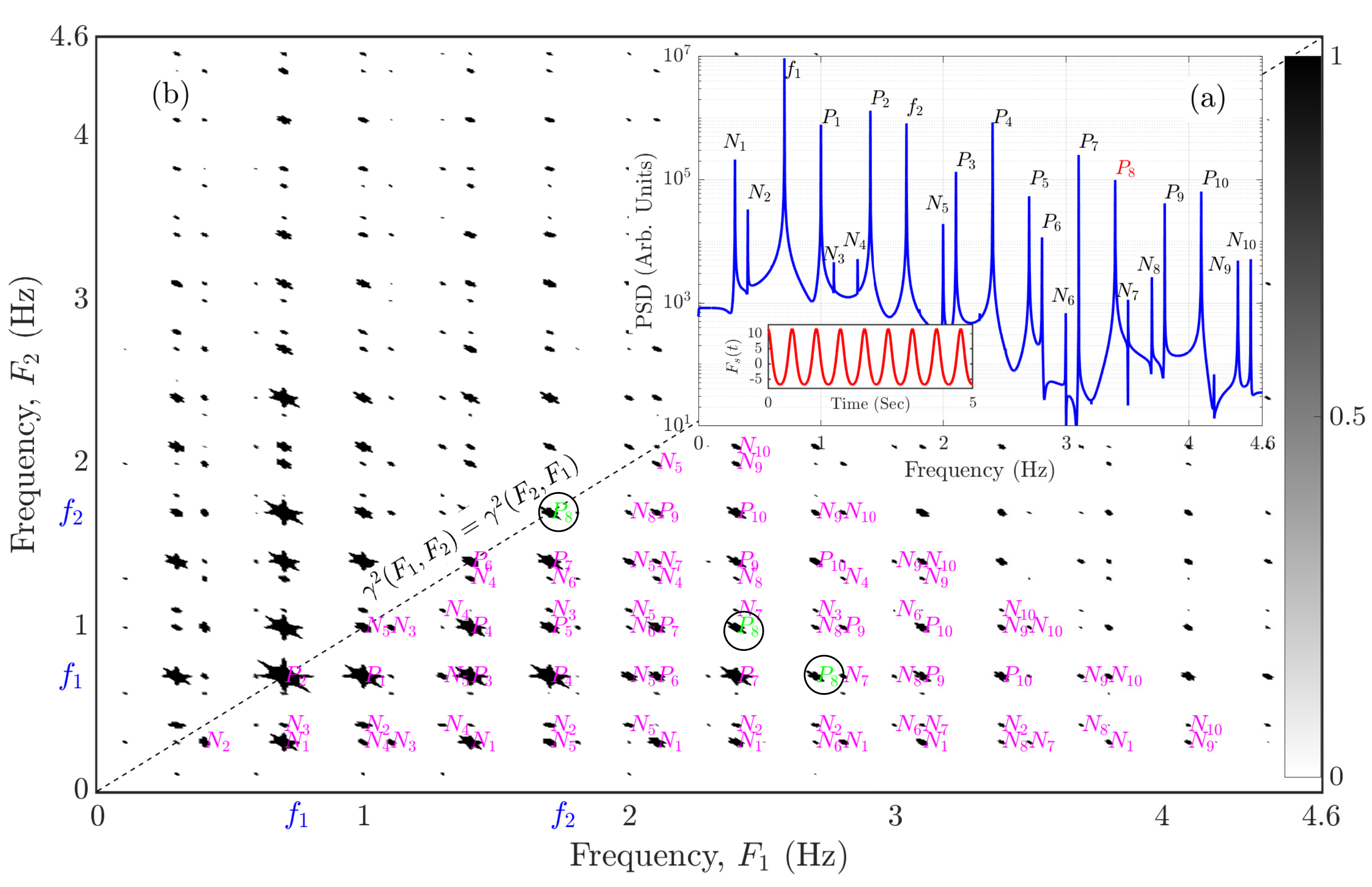}
\caption{
PSD (a) and bicoherence (b) of fKdV for cnoidal-square forcing. 
Inset (a) is the PSD of fKdV for $F_{s}(t) = F_{cn^2}(t) = A_s cn^2[2 K(\kappa_s) f_2 t; \kappa_s]$ with $A_s =18.5$, $f_2=1.7$ Hz and $\kappa_s = 0.9$. Initial parameters are $\mu = 18.5$, $\kappa =0.7$ (such that $f_1 =0.7$ Hz) and  $\alpha = \beta = 1$. 
The small inset within (a) shows the form of cnoidal-square forcing profile.  
Both the PSD (a) and bicoherence (b) show a peak at $P_8 = 2f_2$ (encircled), which is inherent in the cnoidal-square wave forcing.
}
\label{Fig_4}
\end{figure*}
\paragraph*{}
The  normalized bispectrum of a time series~\citep{Kim_IEEE_1979,Nikias_IEEE_1987,Raju_PPCF_2003, Siu_IEEE_2008, Nosenko_PRE_2006} gives the bicoherence and is given by:
\begin{equation}
\label{def_bicohr2}
\gamma^2(F_1, F_2) =\ \frac{|B(F_1, F_2)|^2}{\langle |F(f_1) F(f_2)|^2 \rangle  \langle |F^{\star}(f_1+f_2)|^2 \rangle} .
\end{equation}
Bicoherence gives a measure of phase coherence between the coupled modes. It is a measure of the fraction of power retained by modes due to phase coupling. Theoretically, bicoherence is 1 for phase coupled modes, i.e., modes generated due to coherent nonlinear interaction and 0 for uncoupled modes, i.e., modes generated spontaneously.
\paragraph*{}
Bicoherence is symmetric about the line $F(f_1) = F(f_2)$ because $\gamma^2(F_1,F_2) =\  \gamma^2(F_2,F_1)$. 
For our analysis, we used $M=8192$ sampling data points and $N=16$ time series segments so that the total length of a time series was  $K = M \times N$.
A statistically significant correlation between the coherent modes is determined by the condition $\gamma^2 > \sqrt{6/{2N}} = 0.433$ as discussed  in Siu {\it et al.}~\citep{Siu_IEEE_2008}.
\begin{figure*}[ht!]
\includegraphics[width = 0.99\textwidth]{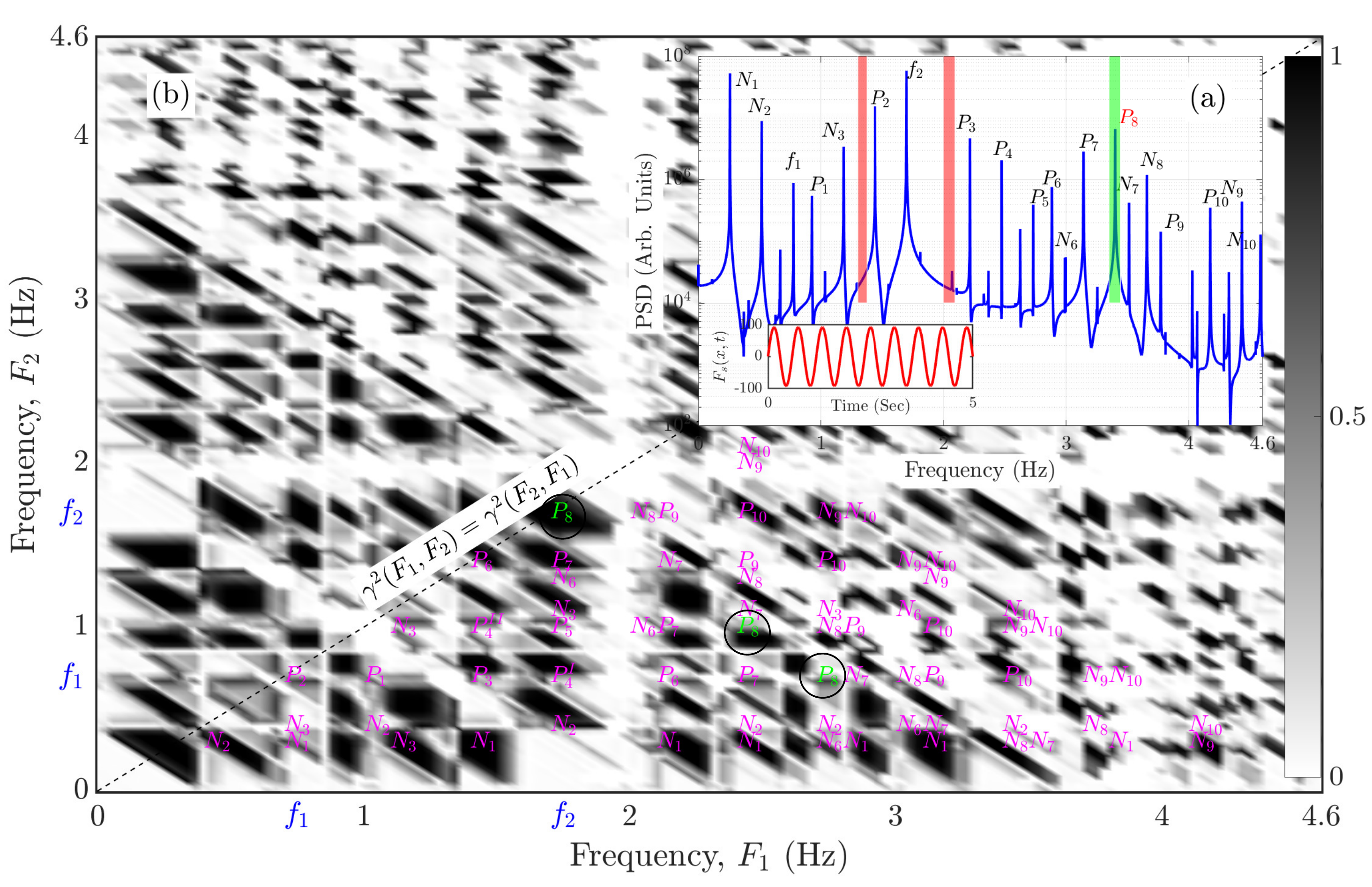}
\caption{
(a) PSD, and (b) bicoherence of fKdV for travelling wave forcing $ F_s(x,t) = A_s \sin(k_s x - 2\pi f_2 t)$ with no initial perturbation, i.e., $n(x,0) = 0$. Here $A_s = 5\mu $, $k_s = 9 k_0$ with $k_0 = (2\pi)/ L$, which corresponds to frequency $f_1 = k_s^3/(2\pi) = 0.7$ Hz for a system of length $L=11\pi$ and $f_2 = 1.7$ Hz. Also $\alpha =\ \beta =\ 1$.
The PSD and the bicoherence show a peak at $P_8 = 2f_2$ (encircled) generated via a coherent nonlinear interaction.}
\label{Fig_5}
\end{figure*}
\subsection{Bispectral analysis for time-dependent sinusoidal forcing}
\label{Bispectral_SIN}
\paragraph*{}
In Fig.~\ref{Fig_3} we show the power spectrum (inset) and the bicoherence of the time series obtained for a sinusoidal forcing in the fKdV model. The frequencies  in the power spectrum as shown in Fig.~\ref{Fig_3}(a) [inset] are listed in column (II) of Table~\ref{Tab_2}. All the listed frequencies match the sum and difference frequencies of $f_1$(fundamental KdV mode), $f_2$ and their harmonics. While the power spectrum confirms the presence of new modes, it does not confirm any frequency or phase coupling. 
\paragraph*{}
To confirm the origin of excited modes, we calculated the bicoherence of the same time series as in Fig.~\ref{Fig_3}(b). Peaks in the power spectrum map to the dark patches in the bicoherence space shown in Fig.~\ref{Fig_3}. For example, $P_4 = 2.4$ Hz in the power spectrum generates a patch ($P_4^{I}$) that is at the intersection of $F_1 = f_1 = 0.7$ Hz and $F_2 = f_2 = 1.7$ Hz confirming a phase coupled and frequency-coupled mode excitation. 
$P_4 = 2.4$ Hz also generates another patch in bicoherence ($P_4^{II}$) at the intersection of $F_1 = 2f_1 = 1.4$ Hz and $F_2 = f_2-f_1 = 1.0$ Hz. 
Table~\ref{Tab_3}, column (IV) lists all possible combinations of frequencies leading to patches in the bicoherence space with $P$ and $N$ nomenclature. 
Peaks marked with $P$ and $N$ are the ones that were observed and missing (not observed), respectively, in the dusty plasma experiment~\citep{Nosenko_PRL_2004} which was used for validating nonlinear mixing in the fKdV model~\citep{Ajaz_POP_2020}.  
Many patches in bicoherence space suggest multiple possible combinations of sum and difference of frequencies leading to a single peak in the power spectrum. Thus the present bispectral analysis confirms the three-wave mixing results for a sinusoidal forcing of the fKdV as presented earlier by Mir {\it et al.}~\citep{Ajaz_POP_2020}.
\paragraph*{}
It should be mentioned that some of the specific patches, marked with a $\star$ in column (IV) of Table~\ref{Tab_3}, do not follow the standard sum rules but are present in the power spectrum as well as in the bicoherence plot. We believe they represent one of the frequencies $F_1$ or $F_2$ that appears on the coordinate axes of Fig.~\ref{Fig_3}. We have independently confirmed their source and presence by constructing a frequency and phase coupled time series generated using all frequencies observed in the power spectrum and switching them on and off to see their impact on the bicoherence plot.
\begin{figure*}[ht!]
\includegraphics[width = 0.9\textwidth]{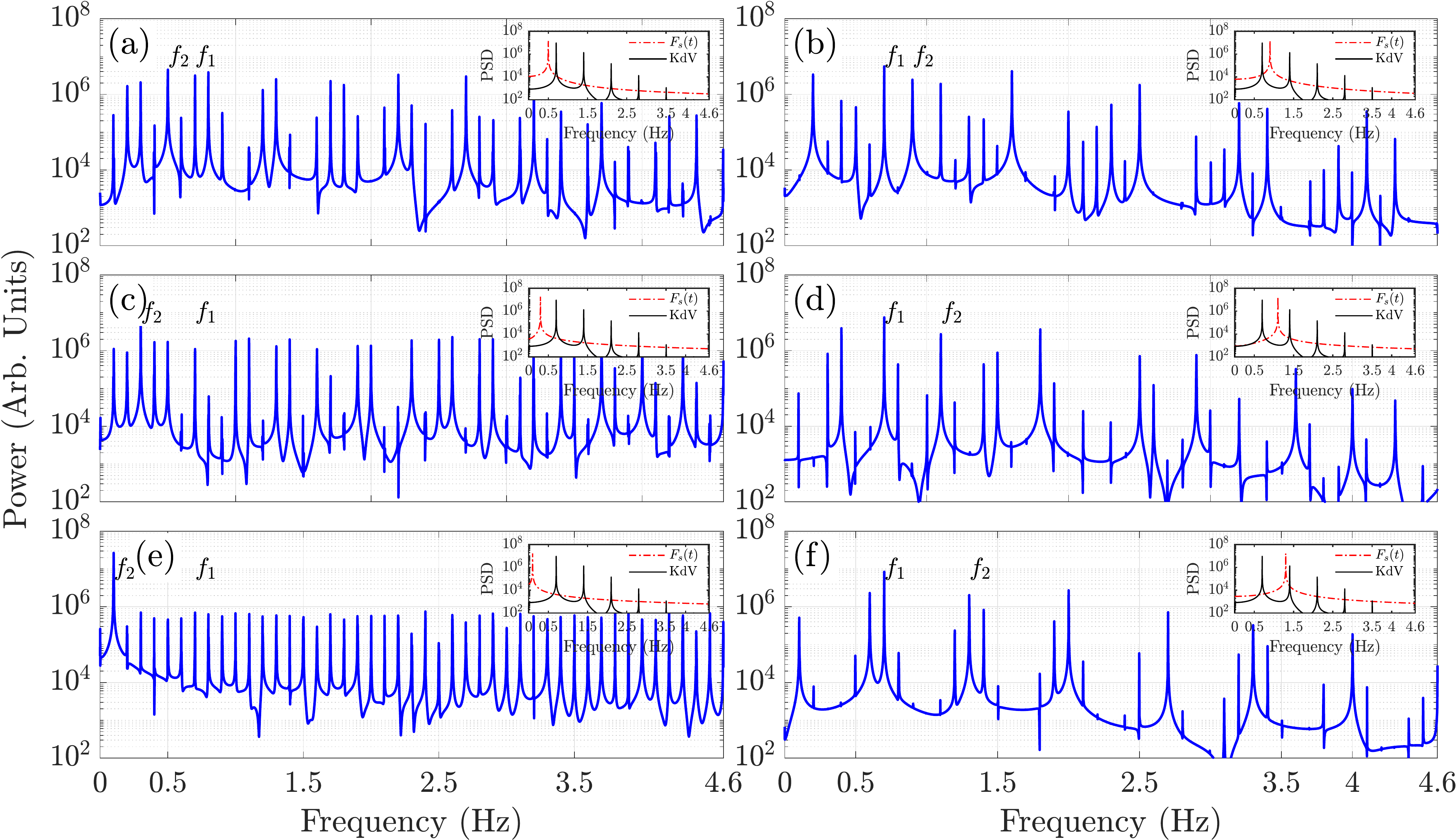}
\caption{
Tailoring the nonlinear mixing profiles of fKdV model with forcing frequency $f_2 < f_1$ (left panel) and $f_2>f_1$ (right panel). 
Left panel:  (a, c, and e) with $f_2 = 0.5$ Hz , $f_2 = 0.3$ Hz and $f_2 = 0.1$ Hz, respectively.
Right panel: (b, d, and f) with $f_2 = 0.9$ Hz , $f_2 = 1.1$ Hz and $f_2 = 1.3$ Hz, respectively. 
Insets in each case display the power spectrum of KdV equation (black bold line) and forcing form (red dash-dotted line) i.e., $F_s(t) = A_s \sin(2\pi f_2 t)$. 
Here $\alpha = \beta = 1$, $\mu = 18.5$, $\kappa = 0.7$ (such that $f_1 = 0.7$ Hz) and $A_s = \mu$ in each case. 
}
\label{Fig_6}
\end{figure*}
\subsection{Bispectral analysis for time-dependent  non-sinusoidal forcing}
\label{Bispectral_CN2}
\paragraph*{}
\textcolor{black}{
In practical scenarios, non-sinusoidal forcing patterns are more probable.  The plasma-based nonlinear dynamical study carried out by Chaubey {\it et al.}~\citep{Neeraj_POP_2015} is one such example, where two nonlinearly excited ion acoustic modes interact with each other. But the studies can easily be generalized for any dispersive media.}
\paragraph*{}
This section provides the bispectral analysis for the time series obtained from the fKdV model with cnoidal and cnoidal-square forms (Fig.~\ref{Fig_4}). Columns III-IV in Table~\ref{Tab_2} list frequencies observed in the power spectra for the two cases as mentioned earlier. Similarly, Columns VI-VII in the Table~\ref{Tab_3} contains all possible frequency combinations due to coherent phase coupling. 
\paragraph*{}
For all non-sinusoidal forcing cases, the general appearance of the power spectra and bicoherence diagrams appear nearly similar to the earlier discussed sinusoidal forcing case within the frequency range of significant amplitudes. An exception is the cnoidal-square case (Fig.~\ref{Fig_4}), where we notice an additional frequency at $P_8 = 2f_2$. Such a  frequency has also been observed in the referred dusty plasma experiment~\citep{Nosenko_PRL_2004}. However, it should be noted that this frequency of  $P_8 = 2f_2$ is also inherently present in the power spectrum of the driver on account of the nature of its profile. So, in this case, it is not possible to unambiguously assert that the appearance of this frequency in the response spectrum is a result of three-wave mixing.
\subsection{Bispectral analysis for travelling wave forcing}
\label{Bispectral_kx_ft}
\paragraph*{}
The bicoherence spectrum of the time series obtained with a travelling wave forcing is shown in Fig.~\ref{Fig_5} with its quantitative values listed in column VII of Table~\ref{Tab_3}.
It shows an absence of the $N_4$ and $N_5$ frequencies and the presence of $P_8 = 2 f_2$  frequency in agreement with the power spectrum analysis. We reiterate that the observation is in contrast to the purely time-dependent sinusoidal forcing and agrees with the results reported in the dusty plasma experiment~\citep{Nosenko_PRL_2004}. 
Also, unlike in the case of a cnoidal-square forcing, the presence of the $P_8 = 2 f_2$ frequency is not an artifact of such a frequency being present in the driver itself. In this case, it is a genuine result of a three-wave mixing process.
\begin{figure*}[ht!]
\includegraphics[width = 0.9\textwidth]{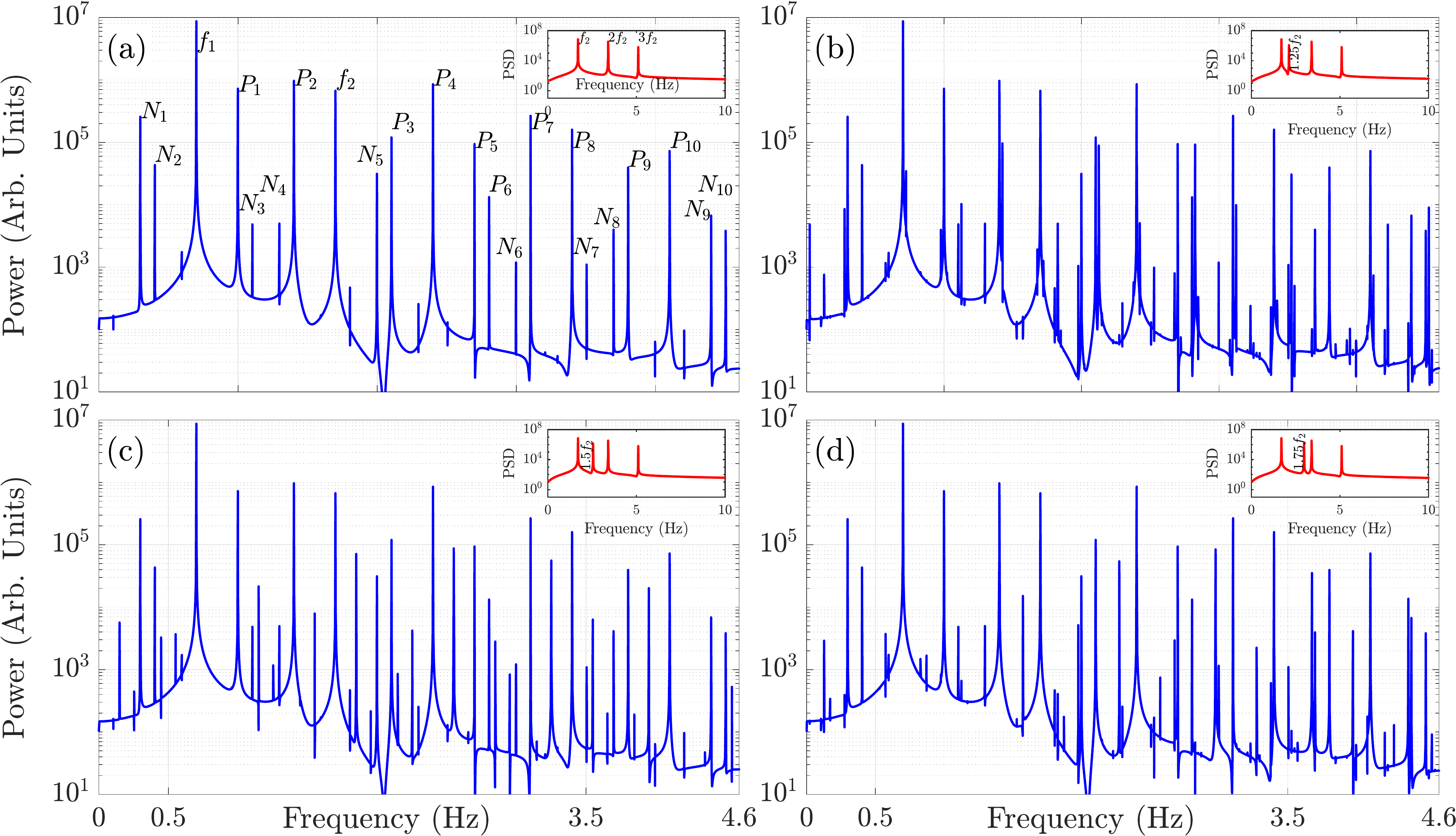}
\caption{
Tailoring the nonlinear mixing profiles by tuning the forcing form.
The approximate form of cnoidal-square forcing as in~\citep{Merlino_POP_2012}
$F_s (t) =\ (A_s/2)\{ \cos(2\pi f_2 t) +\ A_1 \cos(2\pi (2f_2) t) +\ A_2 \cos(2\pi (3f_2) t) +\ A_3 \cos(2\pi (f_T) t)  \}$.
(a) $A_3 = 0$,  
(b) $A_3 = 0.01 A_s$, and $f_T = 1.25f_2$ Hz, 
(c) $A_3 = 0.01 A_s$, and  $f_T = 1.50f_2$ Hz, and 
(d) $A_3 = 0.01 A_s$, and $f_T = 1.75f_2$ Hz.
Here, $A_s = \mu$, $f_2=1.7$ Hz, $A_1 = 0.02 A_s$ and $A_2 = 0.004 A_s$ for each case.
}
\label{Fig_7}
\end{figure*}
\section{Tailoring the mixing pattern through frequency tuning of the driver}
\label{NLM_tunable}
\paragraph*{}
Based on our above-discussed analysis of the influence of the functional form of the driver on the nonlinear mixing process, as seen in the power spectra and bicoherence spectra, we now discuss a possible means of tailoring the nonlinear mixing pattern through the driver parameters. We adopted two approaches for this objective. 
\paragraph*{}
First, we varied the driver frequency $f_2$  towards lower or higher values than the system's natural frequency $f_1$. 
Second, we added specific additional frequency signals to enhance the power in the harmonics of the forcing profiles considered above.
In the first case, we found that when the forcing frequency $f_2 < f_1$, where $f_1$ is the natural frequency of the KdV model, the resultant mixing pattern has a broad spectral form. This is seen in  subplots (a), (c) and (e) of  Fig.~\ref{Fig_6} for cases $f_2 = f_1 - 0.1$, $f_2 = f_1 - 0.3$  and $f_2 = f_1 - 0.5$ respectively. 
On the other hand, when the forcing frequency $f_2 > f_1$, the spectrum of the mixing pattern is sparse and contains fewer frequencies. This is shown in subplots (b), (d) and (f) of the Fig.~\ref{Fig_6} for cases $f_2 = f_1 + 0.1$, $f_2 = f_1 + 0.3$  and $f_2 = f_1 + 0.5$ respectively. 
\paragraph*{}
Such an asymmetry in the nonlinear response of a system to the variation of the driver frequency towards or away from its natural frequency has been observed in other systems such as microwave mixers ~\citep{Pedro_IMD_2003, Lau_APL_1984}. The underlying physical mechanism responsible for this behaviour is a higher-order wave mixing phenomenon called inter-modulation distortion (IMD). Basically the output frequencies of the primary three-wave mixing, namely, $f_1+f_2$, $f_1-f_2$, $2f_1$ and $2f_2$, further mix with the primary modes $f_1$ and $f_2$ to give  rise to $3^{\text{rd}}$ order inter-modulation products such as $2f_1-f_2$ and $2f_2-f_1$. The $3^{\text{rd}}$ order products  may further lead to IMD if they are in proximity of the frequency of the primary modes $f_1$ and $f_2$. This higher-order IMD condition is  better fulfilled  for the condition $f_2 < f_1$ as compared to $f_2 > f_1$.  This results in a broad mixing spectrum for the former case compared to the latter. The additional frequency content created by the $3^{\text{rd}}$ order distortion is also known as ''Spectral Regrowth'' that has not been observed before for the forced KdV model. Controlled regulation of such spectral regrowth through frequency tuning of the driver could prove useful in exploring the nonlinear characteristics of weakly dispersive and small amplitude nonlinear excitations in various plasma and fluid media.
{\renewcommand{\arraystretch}{1.35}
\begin{longtable*}[ht!]{|l|l|l|l|l|l|l|l|l|}
\caption{ \mbox{Phase coherent modes observed in bicoherence for different type of forcing forms.} } 
\label{Tab_3}
\endfirsthead
\endhead
\hline  
Column~(I)    \hspace{0.00cm}   & Column~(II)    \hspace{0.00cm}  & 
Column~(III)  \hspace{0.00cm}   & Column~(IV)    \hspace{0.10cm}  & 
Column~(V)    \hspace{0.00cm}  & Column~(VI)    \hspace{0.00cm}  & 
Column~(VII)  \hspace{0.00cm}  & Column~(VIII)  \hspace{0.00cm}  
\\ \hline 
$F_1$~(Hz)  & $F_2$~(Hz)  & $F_1+F_2$~(Hz)   & Interpretation  
& Sinusoidal  & Cnoidal  & Cnoidal-square  & \textcolor{black}{Travelling wave} \\ \hline
 $f_2-f_1$     & $f_1$       & $f_2-f_1$   & $P_1^{\star}$  
 & \checkmark  & \checkmark  & \checkmark  & \checkmark \\ \hline
 $f_1$         & $f_1$       & $2f_1$   & $P_2$ 
 & \checkmark  & \checkmark  & \checkmark  & \checkmark \\ \hline
 $2f_1$        & $f_1$       & $3f_1$   & $P_3$
 & \checkmark  & \checkmark  & \checkmark  & \checkmark \\ \hline 
 $f_2$         & $f_1$       & $f_1+f_2$   & $P_4$      
 & \checkmark  & \checkmark  & \checkmark  & \checkmark \\ \hline 
 $2f_1$        & $f_2-f_1$   & $f_1+f_2$   & $P_4$  
 & \checkmark  & \checkmark  & \checkmark  & \checkmark \\ \hline
 $f_2$         & $f_2-f_1$   & $2f_2-f_1$  & $P_5$   
 & \checkmark  & \checkmark  & \checkmark  & \checkmark \\ \hline
 $3f_1$        & $f_1$       & $4f_1$   & $P_6$      
 & \checkmark  & \checkmark  & \checkmark  & \checkmark \\ \hline 
 $2f_1$        & $2f_1$      & $4f_1$   & $P_6$    
 & \checkmark  & \checkmark  & \checkmark  & \checkmark \\ \hline
 $f_1+f_2$     & $f_1$       & $2f_1+f_2$  & $P_7$       
 & \checkmark  & \checkmark  & \checkmark  & \checkmark \\ \hline 
 $3f_1$        & $f_2-f_1$   & $2f_1+f_2$  & $P_7$     
 & \checkmark  & \checkmark  & \checkmark  & \checkmark \\ \hline 
 $f_2$         & $2f_1$      & $2f_1+f_2$  & $P_7$    
 & \checkmark  & \checkmark  & \checkmark  & \checkmark \\ \hline 
 $f_2$         & $f_2$       & $2f_2$      & $P_8$    
 &             &             & \checkmark  & \checkmark           \\ \hline                   
 $2f_2-f_1$    & $f_1$       & $2f_2$      & $P_8$       
 &             &             & \checkmark  & \checkmark           \\ \hline 
 $f_2+f_1$     & $f_2-f_1$   & $2f_2$      & $P_8$    
 &             &             & \checkmark  & \checkmark           \\ \hline 
 $2f_1+f_2$    & $f_1$       & $3f_1+f_2$  & $P_9$      
 & \checkmark  & \checkmark  & \checkmark  & \checkmark \\ \hline  
 $4f_1$        & $f_2-f_1$   & $3f_1+f_2$  & $P_9$      
 & \checkmark  & \checkmark  & \checkmark  & \checkmark \\ \hline 
 $f_1+f_2$     & $2f_1$      & $3f_1+f_2$  & $P_9$       
 & \checkmark  & \checkmark  & \checkmark  & \checkmark \\ \hline
 $3f_1$        & $f_2$       & $3f_1+f_2$  & $P_9$     
 & \checkmark  & \checkmark  & \checkmark  & \checkmark \\ \hline 
 $2f_2$        & $f_1$       & $f_1+2f_2$  & $P_{10}$     
 &             &             & \checkmark  & \checkmark \\ \hline      
 $2f_1+f_2$    & $f_2-f_1$   & $f_1+2f_2$  & $P_{10}$     
 & \checkmark  & \checkmark  & \checkmark  & \checkmark \\ \hline   
 $2f_2-f_1$    & $2f_1$      & $f_1+2f_2$  & $P_{10}$     
 & \checkmark  & \checkmark  & \checkmark  & \checkmark \\ \hline 
 $f_1+f_2$     & $f_2$       & $f_1+2f_2$  & $P_{10}$     
 & \checkmark  & \checkmark  & \checkmark  & \checkmark \\ \hline  
 $f_1$         & $f_2-2f_1$   & $f_2-2f_1$  & $N_{1}^{\star}$     
 & \checkmark  & \checkmark   & \checkmark  & \checkmark \\ \hline   
 $2f_1$        & $f_2-2f_1$   & $f_2-2f_1$  & $N_{1}^{\star}$     
 & \checkmark  & \checkmark   & \checkmark  & \checkmark \\ \hline 
 $3f_1$        & $f_2-2f_1$   & $f_2-2f_1$  & $N_{1}^{\star}$     
 & \checkmark  & \checkmark   & \checkmark  & \checkmark \\ \hline 
 $f_2+f_1$     & $f_2-2f_1$   & $f_2-2f_1$  & $N_{1}^{\star}$     
 & \checkmark  & \checkmark   & \checkmark  & \checkmark \\ \hline 
 $4f_1$        & $f_2-2f_1$   & $f_2-2f_1$  & $N_{1}^{\star}$     
 & \checkmark  & \checkmark   & \checkmark  & \checkmark \\ \hline 
 $2f_1+f_2$    & $f_2-2f_1$   & $f_2-2f_1$  & $N_{1}^{\star}$     
 &             &              & \checkmark  & \checkmark \\ \hline
 $3f_1+f_2$    & $f_2-2f_1$   & $f_2-2f_1$  & $N_{1}^{\star}$     
 & \checkmark  & \checkmark   & \checkmark  & \checkmark \\ \hline  
 $3f_1-f_2$    & $f_2-2f_1$   & $3f_1-f_2$  & $N_{2}^{\star}$     
 & \checkmark  & \checkmark   & \checkmark  & \checkmark \\ \hline    
 $f_2-f_1$     & $3f_1-f_2$   & $3f_1-f_2$  & $N_{2}^{\star}$     
 & \checkmark  & \checkmark   & \checkmark  & \checkmark \\ \hline  
 $f_2$         & $3f_1-f_2$   & $3f_1-f_2$  & $N_{2}^{\star}$     
 & \checkmark  & \checkmark   & \checkmark  & \checkmark \\ \hline   
 $f_1+f_2$     & $3f_1-f_2$   & $3f_1-f_2$  & $N_{2}^{\star}$     
 & \checkmark  & \checkmark   & \checkmark  & \checkmark \\ \hline
 $2f_2-f_1$    & $3f_1-f_2$   & $3f_1-f_2$  & $N_{2}^{\star}$     
 & \checkmark  & \checkmark   & \checkmark  & \checkmark \\ \hline 
 $2f_2$        & $3f_1-f_2$   & $3f_1-f_2$  & $N_{2}^{\star}$     
 &             &              & \checkmark  & \checkmark \\ \hline  
 $f_1$         & $3f_1-f_2$   & $4f_1-f_2$  & $N_{3}$     
 & \checkmark  & \checkmark   & \checkmark  & \checkmark \\ \hline                        
 $4f_1-f_2$    & $f_2-2f_1$   & $4f_1-f_2$  & $N_{3}^{\star}$     
 & \checkmark  & \checkmark   & \checkmark  & \checkmark \\ \hline                      
 $4f_1-f_2$    & $f_2-f_1$    & $4f_1-f_2$  & $N_{3}^{\star}$     
 & \checkmark  & \checkmark   & \checkmark  & \checkmark \\ \hline                      
 $f_2$         & $4f_1-f_2$   & $4f_1-f_2$  & $N_{3}^{\star}$     
 & \checkmark  & \checkmark   & \checkmark  & \checkmark \\ \hline                      
 $2f_2-f_1$    & $4f_1-f_2$   & $4f_1-f_2$  & $N_{3}^{\star}$     
 & \checkmark  & \checkmark   & \checkmark  & \checkmark \\ \hline                      
 $f_2-f_1$     & $f_2-2f_1$   & $2f_2-3f_1$  & $N_{4}$     
 & \checkmark  & \checkmark   & \checkmark   &  \\ \hline                      
 $2f_2-3f_1$   & $3f_1-f_2$   & $2f_2-3f_1$  & $N_{4}^{\star}$     
 & \checkmark  & \checkmark   & \checkmark   & \\ \hline             
 $2f_2-3f_1$   & $4f_1-f_2$   & $2f_2-3f_1$  & $N_{4}^{\star}$     
 & \checkmark  & \checkmark   & \checkmark   &  \\ \hline                           
 $2f_1$        & $2f_2-3f_1$  & $2f_2-3f_1$  & $N_{4}^{\star}$     
 & \checkmark  & \checkmark   & \checkmark   &  \\ \hline               
 $3f_1$        & $2f_2-3f_1$  & $2f_2-3f_1$  & $N_{4}^{\star}$     
 &             &              & \checkmark   & \\ \hline 
 $4f_1$        & $2f_2-3f_1$  & $2f_2-3f_1$  & $N_{4}^{\star}$     
 & \checkmark  & \checkmark   & \checkmark   & \\ \hline                                  
 $f_2$        & $f_2-2f_1$   & $2(f_2-f_1)$  & $N_{5}$     
 & \checkmark & \checkmark   & \checkmark    &  \\ \hline  
 $2(f_2-f_1)$ & $3f_1-f_2$   & $2(f_2-f_1)$  & $N_{5}^{\star}$     
 & \checkmark & \checkmark   & \checkmark    &  \\ \hline                              
 $2f_2-3f_1$  & $f_1$        & $2(f_2-f_1)$  & $N_{5}$     
 & \checkmark & \checkmark   & \checkmark    &  \\ \hline                              
 $2(f_2-f_1)$ & $f_1$        & $2(f_2-f_1)$  & $N_{5}^{\star}$     
 & \checkmark & \checkmark   & \checkmark    &  \\ \hline 
 $f_2-f_1$    & $f_2-f_1$    & $2(f_2-f_1)$  & $N_{5}$     
 & \checkmark & \checkmark   & \checkmark    &  \\ \hline                                     
 $2(f_2-f_1)$ & $4f_1-f_2$   & $2(f_2-f_1)$  & $N_{5}^{\star}$     
 & \checkmark & \checkmark   & \checkmark    &  \\ \hline 
 $2(f_2-f_1)$ & $2f_1$       & $2(f_2-f_1)$  & $N_{5}^{\star}$     
 &            &              & \checkmark    &  \\ \hline                                     
 $3f_1$       & $2(f_2-f_1)$ & $2(f_2-f_1)$  & $N_{5}^{\star}$     
 & \checkmark & \checkmark   & \checkmark    &  \\ \hline  
 $2f_2-f_1$   & $f_2-2f_1$   & $3(f_2-f_1)$  & $N_{6}$     
 & \checkmark & \checkmark   & \checkmark    & \checkmark \\ \hline      
 $3(f_2-f_1)$ & $3f_1-f_2$   & $3(f_2-f_1)$  & $N_{6}^{\star}$     
 &            &              & \checkmark    & \checkmark \\ \hline              
 $2(f_2-f_1)$ & $f_2-f_1$    & $3(f_2-f_1)$  & $N_{6}$     
 & \checkmark & \checkmark   & \checkmark    & \checkmark \\ \hline   
 $3(f_2-f_1)$ & $4f_1-f_2$   & $3(f_2-f_1)$  & $N_{6}^{\star}$     
 & \checkmark & \checkmark   & \checkmark    & \checkmark \\ \hline                           
 $f_2$        & $2f_2-3f_1$  & $3(f_2-f_1)$  & $N_{6}$     
 & \checkmark & \checkmark   & \checkmark    & \checkmark \\ \hline    
 $2f_1+f_2$   & $3f_1-f_2$   & $5f_1$        & $N_{7}$     
 & \checkmark & \checkmark   & \checkmark    & \checkmark \\ \hline                      
 $5f_1$       & $f_2-2f_1$   & $5f_1$        & $N_{7}^{\star}$     
 & \checkmark & \checkmark   & \checkmark    & \checkmark \\ \hline                      
 $4f_1$       & $f_1$        & $5f_1$        & $N_{7}$     
 & \checkmark & \checkmark   & \checkmark    & \checkmark \\ \hline                      
 $f_1+f_2$    & $4f_1-f_2$   & $5f_1$        & $N_{7}$     
 & \checkmark & \checkmark   & \checkmark    & \checkmark \\ \hline                      
 $3f_1$       & $2f_1$       & $5f_1$        & $N_{7}$     
 & \checkmark & \checkmark   & \checkmark    & \checkmark \\ \hline                      
 $2f_2$       & $f_2-2f_1$   & $3f_2-2f_1$   & $N_{8}$     
 &            &              & \checkmark    & \checkmark \\ \hline                      
 $3f_2-2f_1$  & $3f_1-f_2$   & $3f_2-2f_1$   & $N_{8}^{\star}$     
 & \checkmark & \checkmark   & \checkmark    & \checkmark \\ \hline                      
 $3(f_2-f_1)$ & $f_1$        & $3f_2-2f_1$   & $N_{8}$     
 & \checkmark & \checkmark   & \checkmark    & \checkmark \\ \hline                      
 $f_1+f_2$    & $2f_2-3f_1$  & $3f_2-2f_1$   & $N_{8}$     
 & \checkmark & \checkmark   & \checkmark    & \checkmark \\ \hline                      
 $2(f_2-f_1)$ & $2f_2-3f_1$  & $3f_2-2f_1$   & $N_{8}$     
 & \checkmark & \checkmark   & \checkmark    & \checkmark \\ \hline                      
 $2(f_2-f_1)$ & $f_2$        & $3f_2-2f_1$   & $N_{8}$     
 & \checkmark & \checkmark   & \checkmark    & \checkmark \\ \hline                      
 $2f_2+f_1$   & $f_2-2f_1$   & $3f_2-f_1$    & $N_{9}$     
 & \checkmark & \checkmark   & \checkmark    & \checkmark \\ \hline                      
 $3f_2-2f_1$  & $f_1$        & $3f_2-f_1$    & $N_{9}$     
 & \checkmark & \checkmark   & \checkmark    & \checkmark \\ \hline  
 $2f_2$       & $f_2-f_1$    & $3f_2-f_1$    & $N_{9}$     
 &            &              & \checkmark    & \checkmark \\ \hline                                  
 $2f_1+f_2$   & $2f_2-3f_1$  & $3f_2-f_1$    & $N_{9}$     
 & \checkmark & \checkmark   & \checkmark    & \checkmark \\ \hline 
 $3(f_2-f_1)$ & $2f_1$       & $3f_2-f_1$    & $N_{9}$     
 & \checkmark & \checkmark   & \checkmark    & \checkmark \\ \hline                                   
 $2f_2-f_1$   & $f_2$        & $3f_2-f_1$    & $N_{9}$     
 & \checkmark & \checkmark   & \checkmark    & \checkmark \\ \hline 
 $f_1+f_2$    & $2(f_2-2f_1)$& $3f_2-f_1$    & $N_{9}$     
 & \checkmark & \checkmark   & \checkmark    & \checkmark \\ \hline
 $2f_2+f_1$   & $3f_1-f_2$   & $4f_1+f_2$    & $N_{10}$     
 & \checkmark & \checkmark   & \checkmark    & \checkmark \\ \hline                               
 $3f_1+f_2$   & $f_1$        & $4f_1+f_2$    & $N_{10}$     
 & \checkmark & \checkmark   & \checkmark    & \checkmark \\ \hline                                    
 $5f_1$       & $f_2-f_1$    & $4f_1+f_2$    & $N_{10}$     
 & \checkmark & \checkmark   & \checkmark    & \checkmark \\ \hline                                 
 $2f_2$       & $4f_1-f_2$   & $4f_1+f_2$    & $N_{10}$     
 &            &              & \checkmark    & \checkmark \\ \hline                                
 $2f_1+f_2$   & $2f_1$       & $4f_1+f_2$    & $N_{10}$     
 & \checkmark & \checkmark   & \checkmark    & \checkmark \\ \hline                                 
 $f_1+f_2$    & $3f_1$       & $4f_1+f_2$    & $N_{10}$     
 & \checkmark & \checkmark   & \checkmark    & \checkmark  \\ \hline
\end{longtable*}
}
\paragraph*{}
We next discuss our second approach towards influencing the nature of the system response spectra by tuning the driver characteristics. As mentioned earlier, we had found that the effects of typical non-sinusoidal (\textit{e.g.}, a cnoidal wave and a cnoidal-square wave) drivers on mixing were very similar to that of a sinusoidal driver (refer to Fig.~\ref{Fig_2}), within a dominant frequency range. The standard driver limits both the amplitude and frequency of their harmonics, restricting the richness of nonlinear mixing. The non-sinusoidal drivers show dominant interactions because of their fundamental mode only, as the amplitudes of their harmonics drop significantly. Hence, in our second approach, we considered tuning the driver by externally pumping up the power in a frequency close to one of the harmonics of the driver. This is done by adding an extra sinusoidal component at a specific frequency and amplitude to the Fourier series representation of a non-sinusoidal forcing term.
\paragraph*{}
Fig.~\ref{Fig_7}(a) shows the time series spectrum obtained from fKdV with a cnoidal-square forcing. The spectrum is similar to the pure sinusoidal forcing case with an additional frequency at $P_8=2f_2$ due to the driver's inherent profile. 
The power at the first harmonic ($2f_2$) is significantly reduced and lies far away from the natural mode ($f_1$). Hence, the fundamental forcing mode ($f_2$) shows more interaction with the natural mode $(f_1)$ and inhibits the interaction of the first harmonic with the natural mode. 
Figs.~\ref{Fig_7}(b), \ref{Fig_7}(c) and \ref{Fig_7}(d) show the spectrum of mixing due to an extra tailoring mode ($f_T$) at $f_T = 1.25 f_2$, $f_T = 1.50 f_2$ and $f_T = 1.75 f_2$ respectively, each with amplitude of $50\% A_{f_2}$, where $A_{f_2}$ is the amplitude of the fundamental mode. This results in a frequency bunching due to the fractional frequency mode ({\it i.e.}, the mode in between the fundamental mode and the first harmonic) close to the fundamental mode leading to an altered frequency spectrum. Thus both these techniques offer one a simple but effective means of tailoring the response spectrum of the driven KdV system. 
\section{Discussion and Conclusions}
\label{summary}
\paragraph*{}
To summarize, we have carried out a detailed numerical study of the driven response of a model KdV equation to a variety of driving sources that range from a simple time-varying sine wave to a spatio-temporally varying plane wave to nonlinear cnoidal waveforms. One of the first objectives of this study is to firmly establish the nature of the process underlying the wave mixing taking place in the system.  An earlier study~\citep{Ajaz_POP_2020}, using a simple time-varying sinusoidal driver, had concluded that three-wave couplings were responsible for the wave mixing in the system. The conclusion was based on identifying some of the combination frequencies in the power spectrum of the nonlinear fluctuations.  However,  the power spectrum information alone is not sufficient to establish the existence of a  three-wave coupling event. This is because the power spectrum does not have the phase coupling information about the interactions. A more precise tool for establishing the existence of three-wave coupling is a bispectral analysis that looks at the triple-correlation of the time series of any dynamical quantity. A finite correlation is obtained for a frequency triad when they are formed by a coherent phase coupling mechanism. In this paper, we have subjected the earlier data of  Ref. \citep{Ajaz_POP_2020} to bispectral analysis and have confirmed that the mixing process arising from a simple time-varying sinusoidal driver is indeed due to three-wave interactions. 
\paragraph*{}
We have next gone on to generalize the findings of Ref. \citep{Ajaz_POP_2020} by changing the nature of the driving source and studying the impact of such changes on the mixing process. In particular, we have chosen time-varying nonlinear drivers in the form of cnoidal waves and the square of cnoidal waves and compared the resultant power spectra and bicoherence spectra with those of the purely sinusoidal driver. The absence of certain spectral lines or the presence of new ones in the response spectra have been identified, and their origin is traced to the changing nature of the natural spectra of the nonlinear driving terms. We have also considered a linear driving term with both a temporal and spatial variation and constitutes a travelling waveform. In this case, we find an additional response frequency that was not present in the case of the sinusoidal driver but had been experimentally observed in Ref. \citep{Nosenko_PRL_2004}. This is not surprising as the sine wave driver, in the earlier model calculation, was adopted as an approximation to the long wavelength ($k \rightarrow 0$) dust acoustic wave that had been excited in the experimental system by an external laser. Such external perturbations can also give rise to a nonlinear excitation {\it e.g.,} as a cnoidal wave that can then act as a driver of the KdV system. Thus both the nonlinear waveforms and the travelling waveform considered as drivers in our model calculation can find useful applications in experimental scenarios. 
\paragraph*{}
Looking at the sensitivity of the response spectra to the nature of the driver, we have further extended our explorations to alter not just the form of the driver but also the frequency of the driver with respect to the natural frequency of the KdV system. Our results point to a novel means of altering the spectral density of the response spectra by manipulating the driver frequency to be smaller or larger than the system's natural frequency or by artificially injecting power in the driver at a  frequency that is somewhat removed from its fundamental frequency. The physical origin of these spectral changes lies in higher-order wave interactions that can create additional frequencies and alter the spectrum's nature. 
\paragraph*{}
Finally, we would like to remark that our results have broad applicability and relevance for understanding nonlinear phenomena in plasmas. Nonlinear mixing is at the heart of wave-wave interactions in plasmas~\citep{Wilhelmsson_ICPIG_1976} and is responsible for such processes as harmonic generation parametric instabilities~\citep{Shukla_PRL_2000}, the onset of weak turbulence, {\it etc.}. Such interactions have been widely studied in the context of laser heating of plasmas~\citep{Brueckner_RMP_1974}, radiofrequency heating in magnetic confinement devices~\citep{Tsytovich_PLA_1973, Ritz_PFB_1989}, and understanding a variety of electrostatic and electromagnetic fluctuations in space plasmas~\citep{Chian_APSC_1996}. One of the principal objectives in such studies is identifying the precise nature of the wave-wave interaction responsible for a particular phenomenon. This becomes a challenging task because many of the nonlinear phenomena often take place simultaneously. Bispectral analysis is a convenient and accurate tool for carrying out such an analysis. The KdV based model used in our analysis is a convenient semi-analytic framework for demonstrating the power and utility of this diagnostic. The conclusions are by no means limited to this model. They are also applicable to any physical scenario where an external (or internally generated) wave drives the plasma and induces nonlinear wave mixing. Our present results can be the basis for understanding nonlinear mixing in plasma systems that are weakly nonlinear and weakly dispersive. Such plasma systems are widely found in space and laboratory setups and have served as convenient media for the study of solitons and other coherent structures. We hope that our work will stimulate further theoretical work in extending the present studies to plasma models of higher dimensions and stronger nonlinearities.

\begin{acknowledgements}
Work done by AM and ST was supported by IIT Jammu Seed Grant No. SG0012. AM and ST acknowledge the use of AGASTYA HPC for present studies.
ST also acknowledges SERB Grant No. CRG/2020/003653 for partial support for the work.
AS is thankful to the Indian National Science Academy (INSA) for their support under the INSA Senior Scientist Fellowship scheme.
\end{acknowledgements}
\bibliography{NLM_DustyPlasma}

\begin{thebibliography}{55}%
\makeatletter
\providecommand \@ifxundefined [1]{%
 \@ifx{#1\undefined}
}%
\providecommand \@ifnum [1]{%
 \ifnum #1\expandafter \@firstoftwo
 \else \expandafter \@secondoftwo
 \fi
}%
\providecommand \@ifx [1]{%
 \ifx #1\expandafter \@firstoftwo
 \else \expandafter \@secondoftwo
 \fi
}%
\providecommand \natexlab [1]{#1}%
\providecommand \enquote  [1]{``#1''}%
\providecommand \bibnamefont  [1]{#1}%
\providecommand \bibfnamefont [1]{#1}%
\providecommand \citenamefont [1]{#1}%
\providecommand \href@noop [0]{\@secondoftwo}%
\providecommand \href [0]{\begingroup \@sanitize@url \@href}%
\providecommand \@href[1]{\@@startlink{#1}\@@href}%
\providecommand \@@href[1]{\endgroup#1\@@endlink}%
\providecommand \@sanitize@url [0]{\catcode `\\12\catcode `\$12\catcode
  `\&12\catcode `\#12\catcode `\^12\catcode `\_12\catcode `\%12\relax}%
\providecommand \@@startlink[1]{}%
\providecommand \@@endlink[0]{}%
\providecommand \url  [0]{\begingroup\@sanitize@url \@url }%
\providecommand \@url [1]{\endgroup\@href {#1}{\urlprefix }}%
\providecommand \urlprefix  [0]{URL }%
\providecommand \Eprint [0]{\href }%
\providecommand \doibase [0]{http://dx.doi.org/}%
\providecommand \selectlanguage [0]{\@gobble}%
\providecommand \bibinfo  [0]{\@secondoftwo}%
\providecommand \bibfield  [0]{\@secondoftwo}%
\providecommand \translation [1]{[#1]}%
\providecommand \BibitemOpen [0]{}%
\providecommand \bibitemStop [0]{}%
\providecommand \bibitemNoStop [0]{.\EOS\space}%
\providecommand \EOS [0]{\spacefactor3000\relax}%
\providecommand \BibitemShut  [1]{\csname bibitem#1\endcsname}%
\let\auto@bib@innerbib\@empty
\bibitem [{\citenamefont {Washimi}\ and\ \citenamefont
  {Taniuti}(1966)}]{Washimi_PRL_1966}%
  \BibitemOpen
  \bibfield  {author} {\bibinfo {author} {\bibfnamefont {H.}~\bibnamefont
  {Washimi}}\ and\ \bibinfo {author} {\bibfnamefont {T.}~\bibnamefont
  {Taniuti}},\ }\href {\doibase 10.1103/PhysRevLett.17.996} {\bibfield
  {journal} {\bibinfo  {journal} {Phys. Rev. Lett.}\ }\textbf {\bibinfo
  {volume} {17}},\ \bibinfo {pages} {996} (\bibinfo {year} {1966})}\BibitemShut
  {NoStop}%
\bibitem [{\citenamefont {Gasch}\ \emph {et~al.}(1986)\citenamefont {Gasch},
  \citenamefont {Berning},\ and\ \citenamefont {J\"ager}}]{Gasch_PRA_1986}%
  \BibitemOpen
  \bibfield  {author} {\bibinfo {author} {\bibfnamefont {A.}~\bibnamefont
  {Gasch}}, \bibinfo {author} {\bibfnamefont {T.}~\bibnamefont {Berning}}, \
  and\ \bibinfo {author} {\bibfnamefont {D.}~\bibnamefont {J\"ager}},\ }\href
  {\doibase 10.1103/PhysRevA.34.4528} {\bibfield  {journal} {\bibinfo
  {journal} {Phys. Rev. A}\ }\textbf {\bibinfo {volume} {34}},\ \bibinfo
  {pages} {4528} (\bibinfo {year} {1986})}\BibitemShut {NoStop}%
\bibitem [{\citenamefont {Hao}\ and\ \citenamefont
  {Maris}(2001)}]{Hao_PRA_2001}%
  \BibitemOpen
  \bibfield  {author} {\bibinfo {author} {\bibfnamefont {H.-Y.}\ \bibnamefont
  {Hao}}\ and\ \bibinfo {author} {\bibfnamefont {H.~J.}\ \bibnamefont
  {Maris}},\ }\href {\doibase 10.1103/PhysRevB.64.064302} {\bibfield  {journal}
  {\bibinfo  {journal} {Phys. Rev. B}\ }\textbf {\bibinfo {volume} {64}},\
  \bibinfo {pages} {064302} (\bibinfo {year} {2001})}\BibitemShut {NoStop}%
\bibitem [{\citenamefont {Leblond}(2008)}]{Leblond_PRA_2008}%
  \BibitemOpen
  \bibfield  {author} {\bibinfo {author} {\bibfnamefont {H.}~\bibnamefont
  {Leblond}},\ }\href {\doibase 10.1103/PhysRevA.78.013807} {\bibfield
  {journal} {\bibinfo  {journal} {Phys. Rev. A}\ }\textbf {\bibinfo {volume}
  {78}},\ \bibinfo {pages} {013807} (\bibinfo {year} {2008})}\BibitemShut
  {NoStop}%
\bibitem [{\citenamefont {Peradze}\ and\ \citenamefont
  {Tsintsadze}(2019)}]{Peradze_LTP_2019}%
  \BibitemOpen
  \bibfield  {author} {\bibinfo {author} {\bibfnamefont {G.~M.}\ \bibnamefont
  {Peradze}}\ and\ \bibinfo {author} {\bibfnamefont {N.~L.}\ \bibnamefont
  {Tsintsadze}},\ }\href {\doibase 10.1063/1.5082319} {\bibfield  {journal}
  {\bibinfo  {journal} {Low Temp. Phys.}\ }\textbf {\bibinfo {volume} {45}},\
  \bibinfo {pages} {103} (\bibinfo {year} {2019})}\BibitemShut {NoStop}%
\bibitem [{\citenamefont {Redor}\ \emph {et~al.}(2019)\citenamefont {Redor},
  \citenamefont {Barth\'elemy}, \citenamefont {Michallet}, \citenamefont
  {Onorato},\ and\ \citenamefont {Mordant}}]{Redor_PRL_2019}%
  \BibitemOpen
  \bibfield  {author} {\bibinfo {author} {\bibfnamefont {I.}~\bibnamefont
  {Redor}}, \bibinfo {author} {\bibfnamefont {E.}~\bibnamefont {Barth\'elemy}},
  \bibinfo {author} {\bibfnamefont {H.}~\bibnamefont {Michallet}}, \bibinfo
  {author} {\bibfnamefont {M.}~\bibnamefont {Onorato}}, \ and\ \bibinfo
  {author} {\bibfnamefont {N.}~\bibnamefont {Mordant}},\ }\href {\doibase
  10.1103/PhysRevLett.122.214502} {\bibfield  {journal} {\bibinfo  {journal}
  {Phys. Rev. Lett.}\ }\textbf {\bibinfo {volume} {122}},\ \bibinfo {pages}
  {214502} (\bibinfo {year} {2019})}\BibitemShut {NoStop}%
\bibitem [{\citenamefont {Congy}\ \emph {et~al.}(2016)\citenamefont {Congy},
  \citenamefont {Kamchatnov},\ and\ \citenamefont {Pavloff}}]{Congy_PRA_2016}%
  \BibitemOpen
  \bibfield  {author} {\bibinfo {author} {\bibfnamefont {T.}~\bibnamefont
  {Congy}}, \bibinfo {author} {\bibfnamefont {A.~M.}\ \bibnamefont
  {Kamchatnov}}, \ and\ \bibinfo {author} {\bibfnamefont {N.}~\bibnamefont
  {Pavloff}},\ }\href {\doibase 10.1103/PhysRevA.93.043613} {\bibfield
  {journal} {\bibinfo  {journal} {Phys. Rev. A}\ }\textbf {\bibinfo {volume}
  {93}},\ \bibinfo {pages} {043613} (\bibinfo {year} {2016})}\BibitemShut
  {NoStop}%
\bibitem [{\citenamefont {Kamchatnov}\ and\ \citenamefont
  {Shchesnovich}(2004)}]{Kamchatnov_PRA_2004}%
  \BibitemOpen
  \bibfield  {author} {\bibinfo {author} {\bibfnamefont {A.~M.}\ \bibnamefont
  {Kamchatnov}}\ and\ \bibinfo {author} {\bibfnamefont {V.~S.}\ \bibnamefont
  {Shchesnovich}},\ }\href {\doibase 10.1103/PhysRevA.70.023604} {\bibfield
  {journal} {\bibinfo  {journal} {Phys. Rev. A}\ }\textbf {\bibinfo {volume}
  {70}},\ \bibinfo {pages} {023604} (\bibinfo {year} {2004})}\BibitemShut
  {NoStop}%
\bibitem [{\citenamefont {Mo}\ \emph {et~al.}(2013)\citenamefont {Mo},
  \citenamefont {Kishek}, \citenamefont {Feldman}, \citenamefont {Haber},
  \citenamefont {Beaudoin}, \citenamefont {O'Shea},\ and\ \citenamefont
  {Thangaraj}}]{Mo_PRL_2013}%
  \BibitemOpen
  \bibfield  {author} {\bibinfo {author} {\bibfnamefont {Y.~C.}\ \bibnamefont
  {Mo}}, \bibinfo {author} {\bibfnamefont {R.~A.}\ \bibnamefont {Kishek}},
  \bibinfo {author} {\bibfnamefont {D.}~\bibnamefont {Feldman}}, \bibinfo
  {author} {\bibfnamefont {I.}~\bibnamefont {Haber}}, \bibinfo {author}
  {\bibfnamefont {B.}~\bibnamefont {Beaudoin}}, \bibinfo {author}
  {\bibfnamefont {P.~G.}\ \bibnamefont {O'Shea}}, \ and\ \bibinfo {author}
  {\bibfnamefont {J.~C.~T.}\ \bibnamefont {Thangaraj}},\ }\href {\doibase
  10.1103/PhysRevLett.110.084802} {\bibfield  {journal} {\bibinfo  {journal}
  {Phys. Rev. Lett.}\ }\textbf {\bibinfo {volume} {110}},\ \bibinfo {pages}
  {084802} (\bibinfo {year} {2013})}\BibitemShut {NoStop}%
\bibitem [{\citenamefont {Bandyopadhyay}\ \emph {et~al.}(2008)\citenamefont
  {Bandyopadhyay}, \citenamefont {Prasad}, \citenamefont {Sen},\ and\
  \citenamefont {Kaw}}]{Bandyopadhyay_PRL_2008}%
  \BibitemOpen
  \bibfield  {author} {\bibinfo {author} {\bibfnamefont {P.}~\bibnamefont
  {Bandyopadhyay}}, \bibinfo {author} {\bibfnamefont {G.}~\bibnamefont
  {Prasad}}, \bibinfo {author} {\bibfnamefont {A.}~\bibnamefont {Sen}}, \ and\
  \bibinfo {author} {\bibfnamefont {P.~K.}\ \bibnamefont {Kaw}},\ }\href
  {\doibase 10.1103/PhysRevLett.101.065006} {\bibfield  {journal} {\bibinfo
  {journal} {Phys. Rev. Lett.}\ }\textbf {\bibinfo {volume} {101}},\ \bibinfo
  {pages} {065006} (\bibinfo {year} {2008})}\BibitemShut {NoStop}%
\bibitem [{\citenamefont {Nakamura}\ and\ \citenamefont
  {Tsukabayashi}(1984)}]{Nakamura_PRL_1984}%
  \BibitemOpen
  \bibfield  {author} {\bibinfo {author} {\bibfnamefont {Y.}~\bibnamefont
  {Nakamura}}\ and\ \bibinfo {author} {\bibfnamefont {I.}~\bibnamefont
  {Tsukabayashi}},\ }\href {\doibase 10.1103/PhysRevLett.52.2356} {\bibfield
  {journal} {\bibinfo  {journal} {Phys. Rev. Lett.}\ }\textbf {\bibinfo
  {volume} {52}},\ \bibinfo {pages} {2356} (\bibinfo {year}
  {1984})}\BibitemShut {NoStop}%
\bibitem [{\citenamefont {Kumar}\ and\ \citenamefont
  {Sen}(2020)}]{Atul_NJP_2020}%
  \BibitemOpen
  \bibfield  {author} {\bibinfo {author} {\bibfnamefont {A.}~\bibnamefont
  {Kumar}}\ and\ \bibinfo {author} {\bibfnamefont {A.}~\bibnamefont {Sen}},\
  }\href {\doibase 10.1088/1367-2630/ab9b6b} {\bibfield  {journal} {\bibinfo
  {journal} {New J. Phys.}\ }\textbf {\bibinfo {volume} {22}},\ \bibinfo
  {pages} {073057} (\bibinfo {year} {2020})}\BibitemShut {NoStop}%
\bibitem [{\citenamefont {Vainberg}\ \emph {et~al.}(1983)\citenamefont
  {Vainberg}, \citenamefont {Meerson},\ and\ \citenamefont
  {Sasorov}}]{Vainberg_RPQE_1983}%
  \BibitemOpen
  \bibfield  {author} {\bibinfo {author} {\bibfnamefont {Y.~R.}\ \bibnamefont
  {Vainberg}}, \bibinfo {author} {\bibfnamefont {B.~I.}\ \bibnamefont
  {Meerson}}, \ and\ \bibinfo {author} {\bibfnamefont {P.~V.}\ \bibnamefont
  {Sasorov}},\ }\href {\doibase doi.org/10.1007/BF01035264} {\bibfield
  {journal} {\bibinfo  {journal} {Radiophys. Quantum Electron.}\ }\textbf
  {\bibinfo {volume} {26}},\ \bibinfo {pages} {1114} (\bibinfo {year}
  {1983})}\BibitemShut {NoStop}%
\bibitem [{\citenamefont {Friedland}\ \emph {et~al.}(2015)\citenamefont
  {Friedland}, \citenamefont {Shagalov},\ and\ \citenamefont
  {Batalov}}]{Friedland_PRE_2015}%
  \BibitemOpen
  \bibfield  {author} {\bibinfo {author} {\bibfnamefont {L.}~\bibnamefont
  {Friedland}}, \bibinfo {author} {\bibfnamefont {A.~G.}\ \bibnamefont
  {Shagalov}}, \ and\ \bibinfo {author} {\bibfnamefont {S.~V.}\ \bibnamefont
  {Batalov}},\ }\href {\doibase 10.1103/PhysRevE.92.042924} {\bibfield
  {journal} {\bibinfo  {journal} {Phys. Rev. E}\ }\textbf {\bibinfo {volume}
  {92}},\ \bibinfo {pages} {042924} (\bibinfo {year} {2015})}\BibitemShut
  {NoStop}%
\bibitem [{\citenamefont {Aranson}\ \emph {et~al.}(1992)\citenamefont
  {Aranson}, \citenamefont {Meerson},\ and\ \citenamefont
  {Tajima}}]{Aranson_PRA_1992}%
  \BibitemOpen
  \bibfield  {author} {\bibinfo {author} {\bibfnamefont {I.}~\bibnamefont
  {Aranson}}, \bibinfo {author} {\bibfnamefont {B.}~\bibnamefont {Meerson}}, \
  and\ \bibinfo {author} {\bibfnamefont {T.}~\bibnamefont {Tajima}},\ }\href
  {\doibase 10.1103/PhysRevA.45.7500} {\bibfield  {journal} {\bibinfo
  {journal} {Phys. Rev. A}\ }\textbf {\bibinfo {volume} {45}},\ \bibinfo
  {pages} {7500} (\bibinfo {year} {1992})}\BibitemShut {NoStop}%
\bibitem [{\citenamefont {Arora}\ \emph {et~al.}(2021)\citenamefont {Arora},
  \citenamefont {Bandyopadhyay}, \citenamefont {Hariprasad},\ and\
  \citenamefont {Sen}}]{Garima_PRE_2021}%
  \BibitemOpen
  \bibfield  {author} {\bibinfo {author} {\bibfnamefont {G.}~\bibnamefont
  {Arora}}, \bibinfo {author} {\bibfnamefont {P.}~\bibnamefont
  {Bandyopadhyay}}, \bibinfo {author} {\bibfnamefont {M.~G.}\ \bibnamefont
  {Hariprasad}}, \ and\ \bibinfo {author} {\bibfnamefont {A.}~\bibnamefont
  {Sen}},\ }\href {\doibase 10.1103/PhysRevE.103.013201} {\bibfield  {journal}
  {\bibinfo  {journal} {Phys. Rev. E}\ }\textbf {\bibinfo {volume} {103}},\
  \bibinfo {pages} {013201} (\bibinfo {year} {2021})}\BibitemShut {NoStop}%
\bibitem [{\citenamefont {Jaiswal}\ \emph {et~al.}(2016)\citenamefont
  {Jaiswal}, \citenamefont {Bandyopadhyay},\ and\ \citenamefont
  {Sen}}]{Surabhi_PRE_2016}%
  \BibitemOpen
  \bibfield  {author} {\bibinfo {author} {\bibfnamefont {S.}~\bibnamefont
  {Jaiswal}}, \bibinfo {author} {\bibfnamefont {P.}~\bibnamefont
  {Bandyopadhyay}}, \ and\ \bibinfo {author} {\bibfnamefont {A.}~\bibnamefont
  {Sen}},\ }\href {\doibase 10.1103/PhysRevE.93.041201} {\bibfield  {journal}
  {\bibinfo  {journal} {Phys. Rev. E}\ }\textbf {\bibinfo {volume} {93}},\
  \bibinfo {pages} {041201} (\bibinfo {year} {2016})}\BibitemShut {NoStop}%
\bibitem [{\citenamefont {Sen}\ \emph {et~al.}(2015)\citenamefont {Sen},
  \citenamefont {Tiwari}, \citenamefont {Mishra},\ and\ \citenamefont
  {Kaw}}]{Sen_ASR_2015}%
  \BibitemOpen
  \bibfield  {author} {\bibinfo {author} {\bibfnamefont {A.}~\bibnamefont
  {Sen}}, \bibinfo {author} {\bibfnamefont {S.}~\bibnamefont {Tiwari}},
  \bibinfo {author} {\bibfnamefont {S.}~\bibnamefont {Mishra}}, \ and\ \bibinfo
  {author} {\bibfnamefont {P.}~\bibnamefont {Kaw}},\ }\href {\doibase
  https://doi.org/10.1016/j.asr.2015.03.021} {\bibfield  {journal} {\bibinfo
  {journal} {Adv. Space Res.}\ }\textbf {\bibinfo {volume} {56}},\ \bibinfo
  {pages} {429 } (\bibinfo {year} {2015})}\BibitemShut {NoStop}%
\bibitem [{\citenamefont {Mir}\ \emph {et~al.}(2020)\citenamefont {Mir},
  \citenamefont {Tiwari}, \citenamefont {Goree}, \citenamefont {Sen},
  \citenamefont {Crabtree},\ and\ \citenamefont {Ganguli}}]{Ajaz_POP_2020}%
  \BibitemOpen
  \bibfield  {author} {\bibinfo {author} {\bibfnamefont {A.~A.}\ \bibnamefont
  {Mir}}, \bibinfo {author} {\bibfnamefont {S.~K.}\ \bibnamefont {Tiwari}},
  \bibinfo {author} {\bibfnamefont {J.}~\bibnamefont {Goree}}, \bibinfo
  {author} {\bibfnamefont {A.}~\bibnamefont {Sen}}, \bibinfo {author}
  {\bibfnamefont {C.}~\bibnamefont {Crabtree}}, \ and\ \bibinfo {author}
  {\bibfnamefont {G.}~\bibnamefont {Ganguli}},\ }\href {\doibase
  10.1063/5.0022482} {\bibfield  {journal} {\bibinfo  {journal} {Phys.
  Plasmas}\ }\textbf {\bibinfo {volume} {27}},\ \bibinfo {pages} {113701}
  (\bibinfo {year} {2020})}\BibitemShut {NoStop}%
\bibitem [{\citenamefont {Nosenko}\ \emph {et~al.}(2004)\citenamefont
  {Nosenko}, \citenamefont {Avinash}, \citenamefont {Goree},\ and\
  \citenamefont {Liu}}]{Nosenko_PRL_2004}%
  \BibitemOpen
  \bibfield  {author} {\bibinfo {author} {\bibfnamefont {V.}~\bibnamefont
  {Nosenko}}, \bibinfo {author} {\bibfnamefont {K.}~\bibnamefont {Avinash}},
  \bibinfo {author} {\bibfnamefont {J.}~\bibnamefont {Goree}}, \ and\ \bibinfo
  {author} {\bibfnamefont {B.}~\bibnamefont {Liu}},\ }\href {\doibase
  10.1103/PhysRevLett.92.085001} {\bibfield  {journal} {\bibinfo  {journal}
  {Phys. Rev. Lett.}\ }\textbf {\bibinfo {volume} {92}},\ \bibinfo {pages}
  {085001} (\bibinfo {year} {2004})}\BibitemShut {NoStop}%
\bibitem [{\citenamefont {Rao}\ and\ \citenamefont
  {Gabr}(1984)}]{Rao_Gabr_SV_1984}%
  \BibitemOpen
  \bibfield  {author} {\bibinfo {author} {\bibfnamefont {T.~S.}\ \bibnamefont
  {Rao}}\ and\ \bibinfo {author} {\bibfnamefont {M.~M.}\ \bibnamefont {Gabr}},\
  }\href {\doibase 10.1007/978-1-4684-6318-7} {\emph {\bibinfo {title} {Lecture
  Notes in Statistics: An Introduction to Bispectral Analysis and Bilinear Time
  Series Models}}}\ (\bibinfo  {publisher} {Springer-Verlag, {New York}},\
  \bibinfo {year} {1984})\BibitemShut {NoStop}%
\bibitem [{\citenamefont {{Nikias}}\ and\ \citenamefont
  {{Raghuveer}}(1987)}]{Nikias_IEEE_1987}%
  \BibitemOpen
  \bibfield  {author} {\bibinfo {author} {\bibfnamefont {C.~L.}\ \bibnamefont
  {{Nikias}}}\ and\ \bibinfo {author} {\bibfnamefont {M.~R.}\ \bibnamefont
  {{Raghuveer}}},\ }\href@noop {} {\bibfield  {journal} {\bibinfo  {journal}
  {Proc. IEEE}\ }\textbf {\bibinfo {volume} {75}},\ \bibinfo {pages} {869}
  (\bibinfo {year} {1987})}\BibitemShut {NoStop}%
\bibitem [{\citenamefont {{Kim}}\ and\ \citenamefont
  {{Powers}}(1979)}]{Kim_IEEE_1979}%
  \BibitemOpen
  \bibfield  {author} {\bibinfo {author} {\bibfnamefont {Y.~C.}\ \bibnamefont
  {{Kim}}}\ and\ \bibinfo {author} {\bibfnamefont {E.~J.}\ \bibnamefont
  {{Powers}}},\ }\href@noop {} {\bibfield  {journal} {\bibinfo  {journal} {IEEE
  Trans. Plasma Sci.}\ }\textbf {\bibinfo {volume} {7}},\ \bibinfo {pages}
  {120} (\bibinfo {year} {1979})}\BibitemShut {NoStop}%
\bibitem [{\citenamefont {van Milligen}\ \emph {et~al.}(1995)\citenamefont {van
  Milligen}, \citenamefont {Hidalgo},\ and\ \citenamefont
  {S\'anchez}}]{Milligen_PRL_1995}%
  \BibitemOpen
  \bibfield  {author} {\bibinfo {author} {\bibfnamefont {B.~P.}\ \bibnamefont
  {van Milligen}}, \bibinfo {author} {\bibfnamefont {C.}~\bibnamefont
  {Hidalgo}}, \ and\ \bibinfo {author} {\bibfnamefont {E.}~\bibnamefont
  {S\'anchez}},\ }\href {\doibase 10.1103/PhysRevLett.74.395} {\bibfield
  {journal} {\bibinfo  {journal} {Phys. Rev. Lett.}\ }\textbf {\bibinfo
  {volume} {74}},\ \bibinfo {pages} {395} (\bibinfo {year} {1995})}\BibitemShut
  {NoStop}%
\bibitem [{\citenamefont {Kim}\ \emph {et~al.}(1997)\citenamefont {Kim},
  \citenamefont {Fonck}, \citenamefont {Durst}, \citenamefont {Fernandez},
  \citenamefont {Terry}, \citenamefont {Paul},\ and\ \citenamefont
  {Zarnstorff}}]{Kim_PRL_1997}%
  \BibitemOpen
  \bibfield  {author} {\bibinfo {author} {\bibfnamefont {J.~S.}\ \bibnamefont
  {Kim}}, \bibinfo {author} {\bibfnamefont {R.~J.}\ \bibnamefont {Fonck}},
  \bibinfo {author} {\bibfnamefont {R.~D.}\ \bibnamefont {Durst}}, \bibinfo
  {author} {\bibfnamefont {E.}~\bibnamefont {Fernandez}}, \bibinfo {author}
  {\bibfnamefont {P.~W.}\ \bibnamefont {Terry}}, \bibinfo {author}
  {\bibfnamefont {S.~F.}\ \bibnamefont {Paul}}, \ and\ \bibinfo {author}
  {\bibfnamefont {M.~C.}\ \bibnamefont {Zarnstorff}},\ }\href {\doibase
  10.1103/PhysRevLett.79.841} {\bibfield  {journal} {\bibinfo  {journal} {Phys.
  Rev. Lett.}\ }\textbf {\bibinfo {volume} {79}},\ \bibinfo {pages} {841}
  (\bibinfo {year} {1997})}\BibitemShut {NoStop}%
\bibitem [{\citenamefont {Nosenko}\ \emph {et~al.}(2006)\citenamefont
  {Nosenko}, \citenamefont {Goree},\ and\ \citenamefont
  {Skiff}}]{Nosenko_PRE_2006}%
  \BibitemOpen
  \bibfield  {author} {\bibinfo {author} {\bibfnamefont {V.}~\bibnamefont
  {Nosenko}}, \bibinfo {author} {\bibfnamefont {J.}~\bibnamefont {Goree}}, \
  and\ \bibinfo {author} {\bibfnamefont {F.}~\bibnamefont {Skiff}},\ }\href
  {\doibase 10.1103/PhysRevE.73.016401} {\bibfield  {journal} {\bibinfo
  {journal} {Phys. Rev. E}\ }\textbf {\bibinfo {volume} {73}},\ \bibinfo
  {pages} {016401} (\bibinfo {year} {2006})}\BibitemShut {NoStop}%
\bibitem [{\citenamefont {{Siu}}\ \emph {et~al.}(2008)\citenamefont {{Siu}},
  \citenamefont {{Ahn}}, \citenamefont {{Ju}}, \citenamefont {{Lee}},
  \citenamefont {{Shin}},\ and\ \citenamefont {{Chon}}}]{Siu_IEEE_2008}%
  \BibitemOpen
  \bibfield  {author} {\bibinfo {author} {\bibfnamefont {K.~L.}\ \bibnamefont
  {{Siu}}}, \bibinfo {author} {\bibfnamefont {J.~M.}\ \bibnamefont {{Ahn}}},
  \bibinfo {author} {\bibfnamefont {K.}~\bibnamefont {{Ju}}}, \bibinfo {author}
  {\bibfnamefont {M.}~\bibnamefont {{Lee}}}, \bibinfo {author} {\bibfnamefont
  {K.}~\bibnamefont {{Shin}}}, \ and\ \bibinfo {author} {\bibfnamefont {K.~H.}\
  \bibnamefont {{Chon}}},\ }\href {\doibase 10.1109/TBME.2007.913418}
  {\bibfield  {journal} {\bibinfo  {journal} {IEEE Trans. Biomedical Eng.}\
  }\textbf {\bibinfo {volume} {55}},\ \bibinfo {pages} {1512} (\bibinfo {year}
  {2008})}\BibitemShut {NoStop}%
\bibitem [{\citenamefont {{Tacchino}}\ \emph {et~al.}(2020)\citenamefont
  {{Tacchino}}, \citenamefont {{Coelli}}, \citenamefont {{Reali}},
  \citenamefont {{Galli}},\ and\ \citenamefont
  {{Bianchi}}}]{Tacchino_IEEE_2020}%
  \BibitemOpen
  \bibfield  {author} {\bibinfo {author} {\bibfnamefont {G.}~\bibnamefont
  {{Tacchino}}}, \bibinfo {author} {\bibfnamefont {S.}~\bibnamefont
  {{Coelli}}}, \bibinfo {author} {\bibfnamefont {P.}~\bibnamefont {{Reali}}},
  \bibinfo {author} {\bibfnamefont {M.}~\bibnamefont {{Galli}}}, \ and\
  \bibinfo {author} {\bibfnamefont {A.~M.}\ \bibnamefont {{Bianchi}}},\ }\href
  {\doibase 10.1109/TBME.2020.2969278} {\bibfield  {journal} {\bibinfo
  {journal} {IEEE Trans. Biomedical Eng.}\ }\textbf {\bibinfo {volume} {67}},\
  \bibinfo {pages} {2696} (\bibinfo {year} {2020})}\BibitemShut {NoStop}%
\bibitem [{\citenamefont {Hillis}\ \emph {et~al.}(2006)\citenamefont {Hillis},
  \citenamefont {Neild}, \citenamefont {Drinkwater},\ and\ \citenamefont
  {Wilcox}}]{Hillis_PRCA_2006}%
  \BibitemOpen
  \bibfield  {author} {\bibinfo {author} {\bibfnamefont {A.~J.}\ \bibnamefont
  {Hillis}}, \bibinfo {author} {\bibfnamefont {S.~A.}\ \bibnamefont {Neild}},
  \bibinfo {author} {\bibfnamefont {B.~W.}\ \bibnamefont {Drinkwater}}, \ and\
  \bibinfo {author} {\bibfnamefont {P.~D.}\ \bibnamefont {Wilcox}},\ }\href
  {\doibase 10.1098/rspa.2005.1620} {\bibfield  {journal} {\bibinfo  {journal}
  {Proc. Math. Phys. Eng. Sci.}\ }\textbf {\bibinfo {volume} {462}},\ \bibinfo
  {pages} {1515} (\bibinfo {year} {2006})}\BibitemShut {NoStop}%
\bibitem [{\citenamefont {{Hall}}\ and\ \citenamefont
  {{Giannakis}}(1995)}]{Hall_IEEE_1995}%
  \BibitemOpen
  \bibfield  {author} {\bibinfo {author} {\bibfnamefont {T.~E.}\ \bibnamefont
  {{Hall}}}\ and\ \bibinfo {author} {\bibfnamefont {G.~B.}\ \bibnamefont
  {{Giannakis}}},\ }\href {\doibase 10.1109/83.392340} {\bibfield  {journal}
  {\bibinfo  {journal} {IEEE Trans. Image Processing}\ }\textbf {\bibinfo
  {volume} {4}},\ \bibinfo {pages} {996} (\bibinfo {year} {1995})}\BibitemShut
  {NoStop}%
\bibitem [{\citenamefont {Flanagan}\ and\ \citenamefont
  {Goree}(2010)}]{Flanagan_POP_2010}%
  \BibitemOpen
  \bibfield  {author} {\bibinfo {author} {\bibfnamefont {T.~M.}\ \bibnamefont
  {Flanagan}}\ and\ \bibinfo {author} {\bibfnamefont {J.}~\bibnamefont
  {Goree}},\ }\href {\doibase 10.1063/1.3524691} {\bibfield  {journal}
  {\bibinfo  {journal} {Phys. Plasmas}\ }\textbf {\bibinfo {volume} {17}},\
  \bibinfo {pages} {123702} (\bibinfo {year} {2010})}\BibitemShut {NoStop}%
\bibitem [{\citenamefont {Ruhunusiri}\ and\ \citenamefont
  {Goree}(2012)}]{Suranga_PRE_2012}%
  \BibitemOpen
  \bibfield  {author} {\bibinfo {author} {\bibfnamefont {W.~D.~S.}\
  \bibnamefont {Ruhunusiri}}\ and\ \bibinfo {author} {\bibfnamefont
  {J.}~\bibnamefont {Goree}},\ }\href {\doibase 10.1103/PhysRevE.85.046401}
  {\bibfield  {journal} {\bibinfo  {journal} {Phys. Rev. E}\ }\textbf {\bibinfo
  {volume} {85}},\ \bibinfo {pages} {046401} (\bibinfo {year}
  {2012})}\BibitemShut {NoStop}%
\bibitem [{\citenamefont {Chaubey}\ \emph {et~al.}(2015)\citenamefont
  {Chaubey}, \citenamefont {Mukherjee}, \citenamefont {Sekar~Iyengar},\ and\
  \citenamefont {Sen}}]{Neeraj_POP_2015}%
  \BibitemOpen
  \bibfield  {author} {\bibinfo {author} {\bibfnamefont {N.}~\bibnamefont
  {Chaubey}}, \bibinfo {author} {\bibfnamefont {S.}~\bibnamefont {Mukherjee}},
  \bibinfo {author} {\bibfnamefont {A.~N.}\ \bibnamefont {Sekar~Iyengar}}, \
  and\ \bibinfo {author} {\bibfnamefont {A.}~\bibnamefont {Sen}},\ }\href
  {\doibase 10.1063/1.4913227} {\bibfield  {journal} {\bibinfo  {journal}
  {Phys. Plasmas}\ }\textbf {\bibinfo {volume} {22}},\ \bibinfo {pages}
  {022312} (\bibinfo {year} {2015})}\BibitemShut {NoStop}%
\bibitem [{\citenamefont {Teng}\ \emph {et~al.}(2009)\citenamefont {Teng},
  \citenamefont {Chang}, \citenamefont {Tseng},\ and\ \citenamefont
  {I}}]{Teng_PRL_2009}%
  \BibitemOpen
  \bibfield  {author} {\bibinfo {author} {\bibfnamefont {L.-W.}\ \bibnamefont
  {Teng}}, \bibinfo {author} {\bibfnamefont {M.-C.}\ \bibnamefont {Chang}},
  \bibinfo {author} {\bibfnamefont {Y.-P.}\ \bibnamefont {Tseng}}, \ and\
  \bibinfo {author} {\bibfnamefont {L.}~\bibnamefont {I}},\ }\href {\doibase
  10.1103/PhysRevLett.103.245005} {\bibfield  {journal} {\bibinfo  {journal}
  {Phys. Rev. Lett.}\ }\textbf {\bibinfo {volume} {103}},\ \bibinfo {pages}
  {245005} (\bibinfo {year} {2009})}\BibitemShut {NoStop}%
\bibitem [{\citenamefont {Mo}\ \emph {et~al.}(2019)\citenamefont {Mo},
  \citenamefont {Singh}, \citenamefont {Raney},\ and\ \citenamefont
  {Purohit}}]{Mo_PRE_2019}%
  \BibitemOpen
  \bibfield  {author} {\bibinfo {author} {\bibfnamefont {C.}~\bibnamefont
  {Mo}}, \bibinfo {author} {\bibfnamefont {J.}~\bibnamefont {Singh}}, \bibinfo
  {author} {\bibfnamefont {J.~R.}\ \bibnamefont {Raney}}, \ and\ \bibinfo
  {author} {\bibfnamefont {P.~K.}\ \bibnamefont {Purohit}},\ }\href {\doibase
  10.1103/PhysRevE.100.013001} {\bibfield  {journal} {\bibinfo  {journal}
  {Phys. Rev. E}\ }\textbf {\bibinfo {volume} {100}},\ \bibinfo {pages}
  {013001} (\bibinfo {year} {2019})}\BibitemShut {NoStop}%
\bibitem [{\citenamefont {Liu}\ \emph {et~al.}(2018)\citenamefont {Liu},
  \citenamefont {Goree}, \citenamefont {Flanagan}, \citenamefont {Sen},
  \citenamefont {Tiwari}, \citenamefont {Ganguli},\ and\ \citenamefont
  {Crabtree}}]{Liu_POP_2018}%
  \BibitemOpen
  \bibfield  {author} {\bibinfo {author} {\bibfnamefont {B.}~\bibnamefont
  {Liu}}, \bibinfo {author} {\bibfnamefont {J.}~\bibnamefont {Goree}}, \bibinfo
  {author} {\bibfnamefont {T.~M.}\ \bibnamefont {Flanagan}}, \bibinfo {author}
  {\bibfnamefont {A.}~\bibnamefont {Sen}}, \bibinfo {author} {\bibfnamefont
  {S.~K.}\ \bibnamefont {Tiwari}}, \bibinfo {author} {\bibfnamefont
  {G.}~\bibnamefont {Ganguli}}, \ and\ \bibinfo {author} {\bibfnamefont
  {C.}~\bibnamefont {Crabtree}},\ }\href {\doibase 10.1063/1.5046402}
  {\bibfield  {journal} {\bibinfo  {journal} {Phys. Plasmas}\ }\textbf
  {\bibinfo {volume} {25}},\ \bibinfo {pages} {113701} (\bibinfo {year}
  {2018})}\BibitemShut {NoStop}%
\bibitem [{\citenamefont {Heinrich}\ \emph {et~al.}(2009)\citenamefont
  {Heinrich}, \citenamefont {Kim},\ and\ \citenamefont
  {Merlino}}]{Heinrich_PRL_2009}%
  \BibitemOpen
  \bibfield  {author} {\bibinfo {author} {\bibfnamefont {J.}~\bibnamefont
  {Heinrich}}, \bibinfo {author} {\bibfnamefont {S.-H.}\ \bibnamefont {Kim}}, \
  and\ \bibinfo {author} {\bibfnamefont {R.~L.}\ \bibnamefont {Merlino}},\
  }\href {\doibase 10.1103/PhysRevLett.103.115002} {\bibfield  {journal}
  {\bibinfo  {journal} {Phys. Rev. Lett.}\ }\textbf {\bibinfo {volume} {103}},\
  \bibinfo {pages} {115002} (\bibinfo {year} {2009})}\BibitemShut {NoStop}%
\bibitem [{\citenamefont {Kholmyansky}\ and\ \citenamefont
  {Gat}(2019)}]{Kholmyansky_PRA_2019}%
  \BibitemOpen
  \bibfield  {author} {\bibinfo {author} {\bibfnamefont {D.}~\bibnamefont
  {Kholmyansky}}\ and\ \bibinfo {author} {\bibfnamefont {O.}~\bibnamefont
  {Gat}},\ }\href {\doibase 10.1103/PhysRevA.100.063809} {\bibfield  {journal}
  {\bibinfo  {journal} {Phys. Rev. A}\ }\textbf {\bibinfo {volume} {100}},\
  \bibinfo {pages} {063809} (\bibinfo {year} {2019})}\BibitemShut {NoStop}%
\bibitem [{\citenamefont {Sharma}\ \emph {et~al.}(2020)\citenamefont {Sharma},
  \citenamefont {Sirse},\ and\ \citenamefont {Turner}}]{Sharma_PSST_2020}%
  \BibitemOpen
  \bibfield  {author} {\bibinfo {author} {\bibfnamefont {S.}~\bibnamefont
  {Sharma}}, \bibinfo {author} {\bibfnamefont {N.}~\bibnamefont {Sirse}}, \
  and\ \bibinfo {author} {\bibfnamefont {M.~M.}\ \bibnamefont {Turner}},\
  }\href {\doibase 10.1088/1361-6595/abbac2} {\bibfield  {journal} {\bibinfo
  {journal} {Plasma Sources Sci. Technol.}\ }\textbf {\bibinfo {volume} {29}},\
  \bibinfo {pages} {114001} (\bibinfo {year} {2020})}\BibitemShut {NoStop}%
\bibitem [{\citenamefont {Qi}\ \emph {et~al.}(2017)\citenamefont {Qi},
  \citenamefont {D'Aguanno},\ and\ \citenamefont {Menyuk}}]{Qi_JOSAB_2017}%
  \BibitemOpen
  \bibfield  {author} {\bibinfo {author} {\bibfnamefont {Z.}~\bibnamefont
  {Qi}}, \bibinfo {author} {\bibfnamefont {G.}~\bibnamefont {D'Aguanno}}, \
  and\ \bibinfo {author} {\bibfnamefont {C.~R.}\ \bibnamefont {Menyuk}},\
  }\href {\doibase 10.1364/JOSAB.34.000785} {\bibfield  {journal} {\bibinfo
  {journal} {J. Opt. Soc. Am. B}\ }\textbf {\bibinfo {volume} {34}},\ \bibinfo
  {pages} {785} (\bibinfo {year} {2017})}\BibitemShut {NoStop}%
\bibitem [{\citenamefont {Farina}\ and\ \citenamefont
  {Bulanov}(2001)}]{Farina_PRL_2001}%
  \BibitemOpen
  \bibfield  {author} {\bibinfo {author} {\bibfnamefont {D.}~\bibnamefont
  {Farina}}\ and\ \bibinfo {author} {\bibfnamefont {S.~V.}\ \bibnamefont
  {Bulanov}},\ }\href {\doibase 10.1103/PhysRevLett.86.5289} {\bibfield
  {journal} {\bibinfo  {journal} {Phys. Rev. Lett.}\ }\textbf {\bibinfo
  {volume} {86}},\ \bibinfo {pages} {5289} (\bibinfo {year}
  {2001})}\BibitemShut {NoStop}%
\bibitem [{\citenamefont {Mahmood}\ and\ \citenamefont
  {Haas}(2014)}]{Mahmood_POP_2014}%
  \BibitemOpen
  \bibfield  {author} {\bibinfo {author} {\bibfnamefont {S.}~\bibnamefont
  {Mahmood}}\ and\ \bibinfo {author} {\bibfnamefont {F.}~\bibnamefont {Haas}},\
  }\href {\doibase 10.1063/1.4899041} {\bibfield  {journal} {\bibinfo
  {journal} {Phys. Plasmas}\ }\textbf {\bibinfo {volume} {21}},\ \bibinfo
  {pages} {102308} (\bibinfo {year} {2014})}\BibitemShut {NoStop}%
\bibitem [{\citenamefont {Flanagan}\ and\ \citenamefont
  {Goree}(2011)}]{Flanagan_POP_2011}%
  \BibitemOpen
  \bibfield  {author} {\bibinfo {author} {\bibfnamefont {T.~M.}\ \bibnamefont
  {Flanagan}}\ and\ \bibinfo {author} {\bibfnamefont {J.}~\bibnamefont
  {Goree}},\ }\href {\doibase 10.1063/1.3544938} {\bibfield  {journal}
  {\bibinfo  {journal} {Phys. Plasmas}\ }\textbf {\bibinfo {volume} {18}},\
  \bibinfo {pages} {013705} (\bibinfo {year} {2011})}\BibitemShut {NoStop}%
\bibitem [{\citenamefont {Salas}(2011)}]{Salas_NLA_2011}%
  \BibitemOpen
  \bibfield  {author} {\bibinfo {author} {\bibfnamefont {A.~H.}\ \bibnamefont
  {Salas}},\ }\href {\doibase https://doi.org/10.1016/j.nonrwa.2010.09.028}
  {\bibfield  {journal} {\bibinfo  {journal} {Nonlinear Anal. Real World
  Appl.}\ }\textbf {\bibinfo {volume} {12}},\ \bibinfo {pages} {1314 }
  (\bibinfo {year} {2011})}\BibitemShut {NoStop}%
\bibitem [{\citenamefont {Abramowitz}\ and\ \citenamefont
  {Stegun}(1965)}]{Abramowitz_Dover_1965}%
  \BibitemOpen
  \bibfield  {author} {\bibinfo {author} {\bibfnamefont {M.}~\bibnamefont
  {Abramowitz}}\ and\ \bibinfo {author} {\bibfnamefont {I.~A.}\ \bibnamefont
  {Stegun}},\ }\href@noop {} {\emph {\bibinfo {title} {Handbook of Mathematical
  Functions : With Formulas, Graphs, and Mathematical Tables}}}\ (\bibinfo
  {publisher} {Dover Publications},\ \bibinfo {address} {New York},\ \bibinfo
  {year} {1965})\BibitemShut {NoStop}%
\bibitem [{\citenamefont {Merlino}\ \emph {et~al.}(2012)\citenamefont
  {Merlino}, \citenamefont {Heinrich}, \citenamefont {Hyun},\ and\
  \citenamefont {Meyer}}]{Merlino_POP_2012}%
  \BibitemOpen
  \bibfield  {author} {\bibinfo {author} {\bibfnamefont {R.~L.}\ \bibnamefont
  {Merlino}}, \bibinfo {author} {\bibfnamefont {J.~R.}\ \bibnamefont
  {Heinrich}}, \bibinfo {author} {\bibfnamefont {S.-H.}\ \bibnamefont {Hyun}},
  \ and\ \bibinfo {author} {\bibfnamefont {J.~K.}\ \bibnamefont {Meyer}},\
  }\href {\doibase 10.1063/1.3693972} {\bibfield  {journal} {\bibinfo
  {journal} {Phys. Plasmas}\ }\textbf {\bibinfo {volume} {19}},\ \bibinfo
  {pages} {057301} (\bibinfo {year} {2012})}\BibitemShut {NoStop}%
\bibitem [{\citenamefont {Raju}\ \emph {et~al.}(2003)\citenamefont {Raju},
  \citenamefont {Sauter},\ and\ \citenamefont {Lister}}]{Raju_PPCF_2003}%
  \BibitemOpen
  \bibfield  {author} {\bibinfo {author} {\bibfnamefont {D.}~\bibnamefont
  {Raju}}, \bibinfo {author} {\bibfnamefont {O.}~\bibnamefont {Sauter}}, \ and\
  \bibinfo {author} {\bibfnamefont {J.~B.}\ \bibnamefont {Lister}},\ }\href
  {\doibase 10.1088/0741-3335/45/4/304} {\bibfield  {journal} {\bibinfo
  {journal} {Plasma Phys. Control. Fusion}\ }\textbf {\bibinfo {volume} {45}},\
  \bibinfo {pages} {369} (\bibinfo {year} {2003})}\BibitemShut {NoStop}%
\bibitem [{\citenamefont {Pedro}\ and\ \citenamefont
  {Carvalho}(2003)}]{Pedro_IMD_2003}%
  \BibitemOpen
  \bibfield  {author} {\bibinfo {author} {\bibfnamefont {J.~C.}\ \bibnamefont
  {Pedro}}\ and\ \bibinfo {author} {\bibfnamefont {N.~B.}\ \bibnamefont
  {Carvalho}},\ }\href@noop {} {\emph {\bibinfo {title} {Intermodulation
  Distortion in Microwave and Wireless Circuits}}}\ (\bibinfo  {publisher}
  {Artech House, {Boston}},\ \bibinfo {year} {2003})\BibitemShut {NoStop}%
\bibitem [{\citenamefont {Lau}\ and\ \citenamefont
  {Yariv}(1984)}]{Lau_APL_1984}%
  \BibitemOpen
  \bibfield  {author} {\bibinfo {author} {\bibfnamefont {K.~Y.}\ \bibnamefont
  {Lau}}\ and\ \bibinfo {author} {\bibfnamefont {A.}~\bibnamefont {Yariv}},\
  }\href {\doibase 10.1063/1.95053} {\bibfield  {journal} {\bibinfo  {journal}
  {Appl. Phys. Lett.}\ }\textbf {\bibinfo {volume} {45}},\ \bibinfo {pages}
  {1034} (\bibinfo {year} {1984})}\BibitemShut {NoStop}%
\bibitem [{\citenamefont {Wilhelmsson}(1976)}]{Wilhelmsson_ICPIG_1976}%
  \BibitemOpen
  \bibfield  {author} {\bibinfo {author} {\bibfnamefont {H.}~\bibnamefont
  {Wilhelmsson}},\ }\href {\doibase
  https://doi.org/10.1016/0378-4363(76)90268-0} {\bibfield  {journal} {\bibinfo
   {journal} {Physica B+C}\ }\textbf {\bibinfo {volume} {82}},\ \bibinfo
  {pages} {52} (\bibinfo {year} {1976})}\BibitemShut {NoStop}%
\bibitem [{\citenamefont {Shukla}(2000)}]{Shukla_PRL_2000}%
  \BibitemOpen
  \bibfield  {author} {\bibinfo {author} {\bibfnamefont {P.~K.}\ \bibnamefont
  {Shukla}},\ }\href {\doibase 10.1103/PhysRevLett.84.5328} {\bibfield
  {journal} {\bibinfo  {journal} {Phys. Rev. Lett.}\ }\textbf {\bibinfo
  {volume} {84}},\ \bibinfo {pages} {5328} (\bibinfo {year}
  {2000})}\BibitemShut {NoStop}%
\bibitem [{\citenamefont {Brueckner}\ and\ \citenamefont
  {Jorna}(1974)}]{Brueckner_RMP_1974}%
  \BibitemOpen
  \bibfield  {author} {\bibinfo {author} {\bibfnamefont {K.~A.}\ \bibnamefont
  {Brueckner}}\ and\ \bibinfo {author} {\bibfnamefont {S.}~\bibnamefont
  {Jorna}},\ }\href {\doibase 10.1103/RevModPhys.46.325} {\bibfield  {journal}
  {\bibinfo  {journal} {Rev. Mod. Phys.}\ }\textbf {\bibinfo {volume} {46}},\
  \bibinfo {pages} {325} (\bibinfo {year} {1974})}\BibitemShut {NoStop}%
\bibitem [{\citenamefont {Tsytovich}\ and\ \citenamefont
  {Stenflo}(1973)}]{Tsytovich_PLA_1973}%
  \BibitemOpen
  \bibfield  {author} {\bibinfo {author} {\bibfnamefont {V.}~\bibnamefont
  {Tsytovich}}\ and\ \bibinfo {author} {\bibfnamefont {L.}~\bibnamefont
  {Stenflo}},\ }\href {\doibase https://doi.org/10.1016/0375-9601(73)90519-7}
  {\bibfield  {journal} {\bibinfo  {journal} {Phys. Lett. A}\ }\textbf
  {\bibinfo {volume} {43}},\ \bibinfo {pages} {7} (\bibinfo {year}
  {1973})}\BibitemShut {NoStop}%
\bibitem [{\citenamefont {Ritz}\ \emph {et~al.}(1989)\citenamefont {Ritz},
  \citenamefont {Powers},\ and\ \citenamefont {Bengtson}}]{Ritz_PFB_1989}%
  \BibitemOpen
  \bibfield  {author} {\bibinfo {author} {\bibfnamefont {C.~P.}\ \bibnamefont
  {Ritz}}, \bibinfo {author} {\bibfnamefont {E.~J.}\ \bibnamefont {Powers}}, \
  and\ \bibinfo {author} {\bibfnamefont {R.~D.}\ \bibnamefont {Bengtson}},\
  }\href {\doibase 10.1063/1.859082} {\bibfield  {journal} {\bibinfo  {journal}
  {Phys. Fluids B: Plasma Phys.}\ }\textbf {\bibinfo {volume} {1}},\ \bibinfo
  {pages} {153} (\bibinfo {year} {1989})}\BibitemShut {NoStop}%
\bibitem [{\citenamefont {Chian}(1996)}]{Chian_APSC_1996}%
  \BibitemOpen
  \bibfield  {author} {\bibinfo {author} {\bibfnamefont {A.-L.}\ \bibnamefont
  {Chian}},\ }\href@noop {} {\bibfield  {journal} {\bibinfo  {journal}
  {Astrophys. Space Sci.}\ }\textbf {\bibinfo {volume} {242}},\ \bibinfo
  {pages} {249} (\bibinfo {year} {1996})}\BibitemShut {NoStop}%
\end{thebibliography}%
\end{document}